\documentclass[twocolumn,showpacs,nofootinbib]{revtex4-1}
\usepackage{graphicx,amsmath}
\usepackage{epstopdf}
\usepackage{hyperref}

\begin{document}

\title{Testing statics-dynamics equivalence at the spin-glass transition in three dimensions}
\author{Luis Antonio Fern\'andez}
\affiliation{Departamento de F\'isica Te\'orica I, Universidad Complutense, 28040 Madrid, Spain}
\affiliation{Instituto de Biocomputaci\'on y F\'isica de Sistemas Complejos (BIFI), Zaragoza, Spain}
\author{V\'{\i}ctor Mart\'{\i}n-Mayor}
\affiliation{Departamento de F\'isica Te\'orica I, Universidad Complutense, 28040 Madrid, Spain}
\affiliation{Instituto de Biocomputaci\'on y F\'isica de Sistemas Complejos (BIFI), Zaragoza, Spain}
\date{\today}

\begin{abstract}
The statics-dynamics correspondence in spin glasses relate non-equilibrium
results on large samples (the experimental realm) with equilibrium quantities
computed on small systems (the typical arena for theoretical
computations). Here we employ statics-dynamics equivalence to study the Ising
spin-glass critical behavior in three dimensions. By means of Monte Carlo
simulation, we follow the growth of the coherence length (the size of the
glassy domains), on lattices too large to be thermalized. Thanks to the large
coherence lengths we reach, we are able to obtain accurate results in
excellent agreement with the best available equilibrium computations. To do
so, we need to clarify the several physical meanings of the dynamic exponent
close to the critical temperature.
\end{abstract}
 \maketitle

The glass transition, the dramatic dynamic slowdown experienced by
spin-glasses, fragile molecular glasses, polymers, colloids, etc., upon
approaching their glass temperature $T_\mathrm{g}$, has long puzzled
scientists~\cite{cavagna:09} The phenomenon has been long suspected to be
caused by the growth of a characteristic length~\cite{adam:65}, an issue under
current investigation~\cite{weeks:00,berthier:05,gutierrez:14}.

Spin-glasses enjoy a privileged status in this context, for a number of
reasons. \emph{First,} their glass transition is a \emph{bona fide} phase
transition at
$T_\mathrm{c}=T_\mathrm{g}$~\cite{gunnarsson:91,palassini:99,ballesteros:00}. \emph{Second,}
consider a rapid quench from high-temperature to the working temperature
$T<T_\mathrm{c}$, where the system is left to equilibrate for a time
$t_\mathrm{w}$. The system remains perennially out equilibrium. This
\emph{aging} process~\cite{vincent:97} consists in the growth of glassy
magnetic domains (which reminds coarsening~\cite{bray:94}).  The size of these
domains $\xi(t_\mathrm{w})$ is experimentally accesible, and it can be as
large as 100 lattice spacings~\cite{joh:99,bert:04} (enormously larger than
any length scale identified on molecular
liquids~\cite{berthier:05,gutierrez:14}).\emph{Third,} the growth of
$\xi(t_\mathrm{w})$ has been well studied
numerically~\cite{rieger:93,kisker:96,marinari:00,marinari:00b,berthier:02,berthier:05b,jaubert:07,janus:08b,janus:09b,manssen:14}. In
particular, the dedicated Janus computer~\cite{janus:08} has allowed to cover
$t_\mathrm{w}$ ranging from picoseconds to 0.1
seconds~\cite{janus:08b,janus:09b}.\emph{Fourth}, a statics-dynamics
correspondence is expected~\cite{franz:98}: detailed dictionaries have been
built~\cite{barrat:01,janus:10b}, relating equilibrium results on finite
systems (the typical setting for numerical simulations) with non-equilibrium
results on macroscopic (or mesoscopic) samples.

The statics-dynamics equivalence is particularly exciting, because it brings the
much awaited possibility of detailed comparisons between experimental results
and theoretical computations. In fact, experimental effort has been recently
devoted to the measurement of $\xi(t_\mathrm{w})$ with that
end~\cite{joh:99,jonsson:02b,bert:04,nakamae:12,guchhait:14}. Unfortunately,
appealing as it is, the static-dynamic equivalence has not yet produced new
insights (in fact, not even the mutual consistency of different
dictionaries~\cite{barrat:01,janus:10b} has been shown).

Here, we obtain a complete characterization of the critical behavior
of the three-dimensional Ising spin-glass based solely on the
statics-dynamics equivalence. Our Monte Carlo simulations follow the
growth of $\xi(t_\mathrm{w})$ on lattices too large to be
equilibrated. In this way, we obtain the largest coherence lengths
ever obtained in a simulation (up to 25 lattice spacing). Thus armed,
we obtain fairly accurate estimates of the critical exponents. Our
results are completely consistent with the best~\emph{equilibrium}
computations on small lattices~\cite{hasenbusch:08,janus:13}. Our
analysis is obviously related to dynamic scaling~\cite{ozeki:07}, with
an important difference. We find it mandatory to eliminate time, in
favour of the coherence length $\xi(t_\mathrm{w})$. The reason,
explained below, is in that the dynamic exponent $z$ changes its
physical meaning at $T_\mathrm{c}$. Last, but not least, we show in
Appendix~\ref{sect:MSC} how to perform on conventional processors
investigations previously regarded as impossible without special
computers.

The Hamiltonian for the $D\!=\!3$ Edwards-Anderson model with nearest-neighbors interactions is
\begin{equation}
{\cal H}=-\sum_{\langle \boldsymbol{x},\boldsymbol{y}\rangle} J_{\boldsymbol{x},\boldsymbol{y}}\, \sigma_{\boldsymbol{x}}\, \sigma_{\boldsymbol{y}}\,.\label{EA-H}
\end{equation}
The spins $\sigma_{\boldsymbol{x}}=\pm 1$ are placed on the nodes,
$\boldsymbol{x}$, of a cubic lattice of linear size $L=256$ and
periodic boundary conditions.  The couplings
$J_{\boldsymbol{x},\boldsymbol{y}}=\pm 1$ are chosen randomly with
$50\%$ probability, and are quenched variables. Each coupling choice is named
a \emph {sample}. We
denote by $\overline{(\cdot \cdot \cdot)}$ the average over the
couplings.  Model (\ref{EA-H}) undergoes a spin-glass transition at
$T_\mathrm{c}=1.1019(29)$~\cite{janus:13}.

\begin{figure}
\begin{center}
\includegraphics[angle=270,width=\columnwidth]{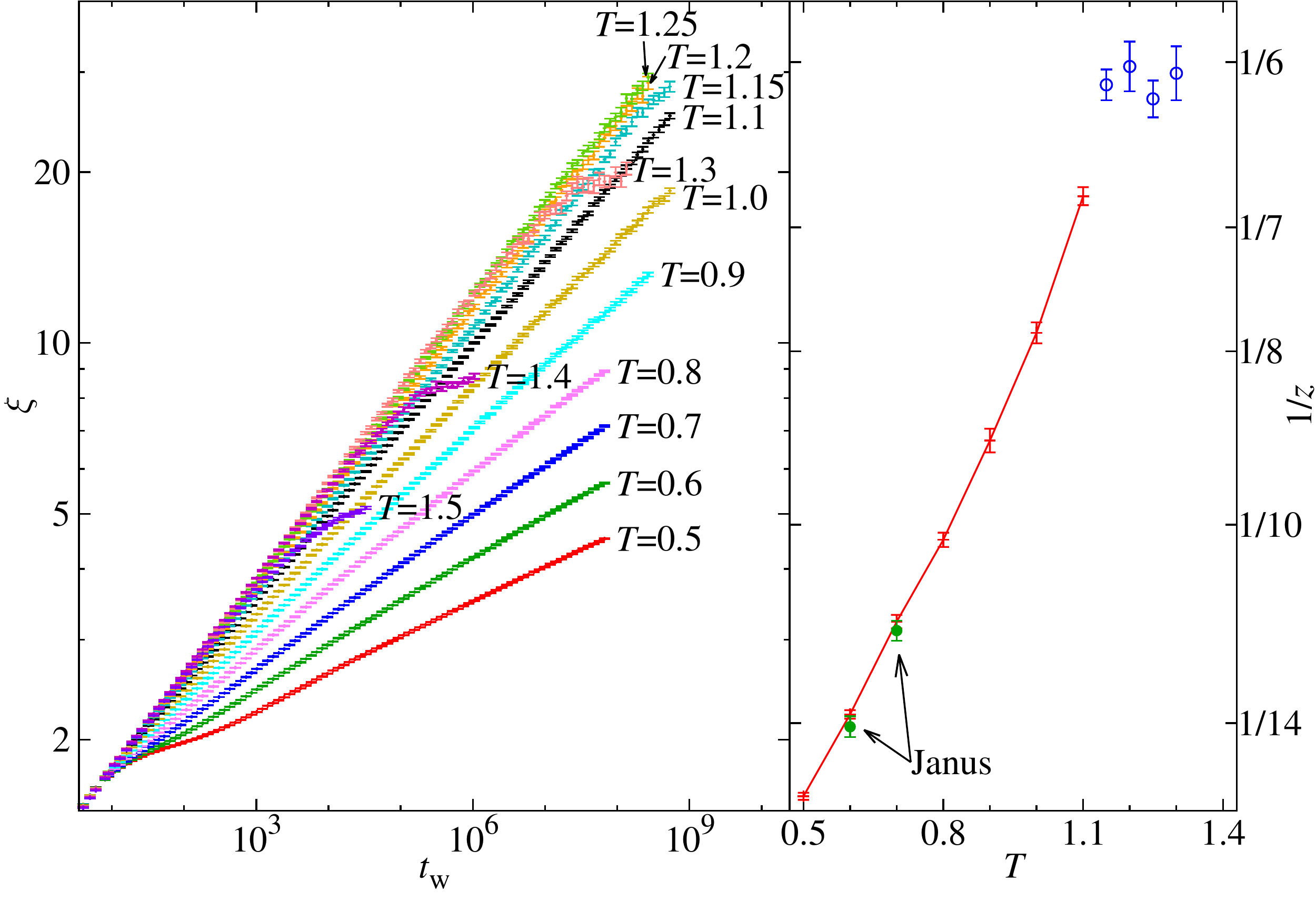}
\caption{(Color online) {\bf Left:} Coherence length
  $\xi_{12}(t_\mathrm{w})$ vs. Monte Carlo time, as computed for
  model~\eqref{EA-H} on lattices of size $L=256$, for several
  temperatures [$T_\mathrm{c}=1.1019(29)$~\cite{janus:13}]. One Monte
  Carlo step corresponds to 1 picosecond in physical
  time~\cite{mydosh:93}.  For $T\geq 1.3$, we reach equilibrium. {\bf
    Right:} Dynamic exponent $z(T)$ computed in the non-equilibrium
  regime $\xi(t_\mathrm{w})\sim t_\mathrm{w}^{1/z(T)}$. Joined red
  points stand for $T\leq T_\mathrm{c}$.  Note the constant value
  $z(T> T_\mathrm{c})\approx 6$ (blue circles).  We perform the fits
  for $t_\mathrm{w}>2^{20}\approx 10^6$ Monte Carlo steps (but for
  $T=1.3$ where $ 2^{16} \leq t_\mathrm{w}\leq 2^{20}$, in order to
  avoid thermalization).  We also show Janus data~\cite{janus:09b}
  (green circles), computed for longer times. We only show the $L=80$
  Janus data at temperatures free from finite-size artifacts.}
\label{fig:xi_tw}
\vspace{-0.7cm}
\end{center}
\end{figure}

We study the direct quench, the simplest dynamic protocol.  At the
starting time $t_\mathrm{w}=0$, the system is in a random
configuration (i.e. $T=\infty$). We place it instantaneously at the
working temperature $T$ and follow the evolution as $t_\mathrm{w}$
increases, Fig.~\ref{fig:xi_tw}. Our time unit is the Monte Carlo step
(a full lattice Metropolis sweep).\footnote{We have simulated the same
  50 samples at $T=1.5$ ($t_\mathrm{w}\leq 2^{23}$), $T=1.4$
  ($t_\mathrm{w}\leq 2^{25}$), $T=0.5,0.6,0.7,0.8$ ($t_\mathrm{w}\leq
  2^{26}$), $T=0.9,1.2,1.25.1.3$ ($t_\mathrm{w}\leq 2^{28}$), and at
  $T=1.0,1.1,1.15$ ($t_\mathrm{w}\leq 2^{29}$). For each sample, we
  simulate four independent systems (replicas),
  $\{\sigma_{\boldsymbol{x}}^{(a)}\}$ $a=1,\ldots,4$ [8 replicas at
    $T=1.1\approx T_\mathrm{c}$, and (to control the possibility of
    thermalization effects) at $T=1.25$].}

Metropolis dynamics belongs to the Universality Class of the physical
evolution (it is an instance of the so-called model A
dynamics~\cite{hohenberg:77}), and is straightforward to
implement~\cite{landau:05}. However, our aim is to reach large $L$ and
$t_\mathrm{w}$. Rather than resorting to special
hardware~\cite{janus:08,manssen:14,lulli:14,fang:14,feng:14}, we
employ synchronous multi-spin coding on standard CPUs. In a naive
implementation random number generation is a major cost. However, our
minimal energy barrier is 4, rarely overcome at the temperatures of
interest [for instance, exp(-4/$T_\mathrm{c}$)$\approx$ 0.026]. Hence,
the Gillespie method~\cite{gillespie:77,bortz:75} allows for major
savings (see Appendix \ref{sect:MSC}).\footnote{Also, we employ {\em
    Pthreads\/} to simulate a single system in multicore
  processors. Our best timings for $L=256$ at $T_\mathrm{c}$ are: (a)
  An 8-core Intel(R) Xeon(R) CPU E5-2690: a 8-threads simulation of a
  single system at 50 ps/spin-flip; (b) A single 16-core AMD Opteron
  (TM) 6272 processor: a 16-threads simulation of a single system at
  62 ps/spin-flip. For comparison, a single Janus FPGA runs two $L=80$
  systems at 32 ps/spin-flip each~\cite{janus:08,janus:08b}.}

We compute the coherence-length from the correlation function
of the replica field $q_{\boldsymbol{x}}(t_\mathrm{w})\equiv \sigma_{\boldsymbol x}^{(a)}(t_\mathrm{w}) \sigma_{\boldsymbol{x}}^{(b)}(t_\mathrm{w})\,$:\footnote{Having 4 replicas at our disposal (8 replicas for $T=1.1$, $1.25$) we average over the 6 (28) possible pairings of replica indices.}
\begin{equation}
C_4(\boldsymbol r,t_\mathrm{w})=\overline{L^{-3}\sum_{\boldsymbol x} q_{\boldsymbol x}(t_\mathrm{w}) q_{\boldsymbol x+\boldsymbol r}(t_\mathrm{w})}\,.\label{C4DEF}
\end{equation}
We restrict the displacement $\boldsymbol r$ to a lattice axis and
compute integrals $I_k(t_\mathrm{w})=\int_0^\infty
\mathrm{d}\,r\,r^kC_4(r,t_\mathrm{w})\,$. Then,
$\xi_{1,2}(t_\mathrm{w})=I_2(t_\mathrm{w})/I_1(t_\mathrm{w})$~\cite{janus:08b,janus:09b}. In
all cases, we find $L> 10\, \xi_{1,2}(t_\mathrm{w})$ hence we regard
our data as representative of the thermodynamic
limit~\cite{janus:08b}.

Fig.~\ref{fig:xi_tw} shows a rather accurate algebraic growth,
$\xi(t_\mathrm{w})\sim
t_\mathrm{w}^{1/z(T)}$~\cite{joh:99,marinari:00}.\footnote{Other
  laws~\cite{bouchaud:01} are numerically indistinguishable from a
  power.} Yet, there is some controversy.  On the one hand,
low-temperature data suggest $z(T\leq T_\mathrm{c}) \approx
z_\mathrm{c}
T_\mathrm{c}/T$~\cite{joh:99,marinari:00,janus:08b,janus:09b}.  On the
other hand, in Ref.~\cite{liu:14} a temperature varying protocol with
$T\geq T_\mathrm{c}$ produced a numerical value
[$z_\mathrm{c}=5.85(9)$ for $J=\pm 1$ or $z_\mathrm{c}=6.00(19)$ for
  Gaussian couplings] hardly consistent with the low-temperature
$z_\mathrm{c}=6.86(16)$~\cite{janus:08b,janus:09b}.

Our own data, Fig.~\ref{fig:xi_tw}--right, suggest an exponent $z(T)$
discontinuous at $T_\mathrm{c}$.  Of course, this might be an effect
of our $z(T)$ being an effective exponent (due to our fitting
time-window). But this is not a logical necessity.

Indeed, exponent $z(T)$ carries different meanings. For $T<
T_\mathrm{c}$ it describes (glassy) coarsening: the coherence length
grows forever as $\xi(t_\mathrm{w})\sim t_\mathrm{w}^{1/z(T)}$. Yet,
$z(T>T_\mathrm{c})$ is concerned with equilibration. One has a
characteristic time $\tau(T)$ (when $\xi(\tau,T)$ reaches, say, $90\%$
of the equilibrium $\xi_{\mathrm{eq}}$) and then $\tau(T)\propto
[\xi_\mathrm{eq}(T)]^{z^*}$. In fact, for the simplest non-trivial
model (the $D=2$ Ising ferromagnet) the coarsening exponent is
$z_{\mathrm{FM}}(T<T_{\mathrm{c}}^{{\mathrm{FM}}})=2$~\cite{bray:94},
while $z^*_{\mathrm{c},\mathrm{FM}}=2.1667(5)$~\cite{nightingale:00} for critical
equilibration.

Clearly, this delicate cross-over will require further
investigation. Yet, we have rationalized why a $T\geq T_\mathrm{c}$
protocol~\cite{liu:14} produces $z(T>T_\mathrm{c})\approx 6$.

These complications reinforce our choice of basing finite-time scaling
on $\xi_{1,2}(t_\mathrm{w})$, rather than on
$t_\mathrm{w}$~\cite{nakamura:03,ozeki:07,roma:13}. To do so, we adapt
Binder's method~\cite{binder:81} (in  Appendix \ref{sect:I2}  we explore
another possibility~\cite{nakamura:10} that turns out to be less
accurate). Let $q(B_l,t_\mathrm{w}) = \sum_{_{\boldsymbol x}\in B_l}
q_{\boldsymbol x}(t_\mathrm{w})/l^3$ be the average of the replica
field on a cubic box of side $l$. We compute
$q_k(l,t_\mathrm{w})=\overline{q^k(B_l,t_\mathrm{w})}$, 
its $k$-th power averaged over samples, replica parings, as well as over boxes
$B_l$.  Binder's ratio
$U_4(l,t_\mathrm{w},T)=\overline{q_4(l,t_\mathrm{w})}/\overline{q_2(l,t_\mathrm{w})}^2$
is a dimensionless parameter likely to display Universal behavior (for
instance, $U_4(l,t_\mathrm{w},T)\to 3$ when $l\gg \xi(t_\mathrm{w})$
due to the Central Limit Theorem, see also Ref.~\cite{marinari:96}).

The analogy with Finite Size Scaling impels us to change variables: 
$y=[T-T_\mathrm{c}][\xi(t_\mathrm{w},T)]^{1/\nu}$ and
$\lambda=l/\xi(t_\mathrm{w},T)$. Then, barring subleading corrections to
scaling, we expect:
\begin{equation}\label{eq:fts}
U_4(l,t_\mathrm{w},T)=f(y,\lambda)+[\xi(t_\mathrm{w})]^{-\omega} g(y,\lambda),
\end{equation}
where $\nu$ is the correlation-length critical exponent, $\omega$ is
the leading corrections to scaling exponent, and $f$ and $g$ are
dimensionless scaling functions. Note that the independent variables
in the l.h.s. of Eq.~\eqref{eq:fts} ($l$, $t_\mathrm{w}$ and $T$) are
discrete. Yet, the r.h.s. variables ($\xi,y$ and $\lambda$) are
continuous.  We solve this problem by means of polynomial
interpolations (see Appendix \ref{sect:polynomials}). Errors are
estimated with the jackknife method~\cite{amit:05}, computed over the
samples.

\begin{figure}
\begin{center}
\includegraphics[angle=270,width=\columnwidth]{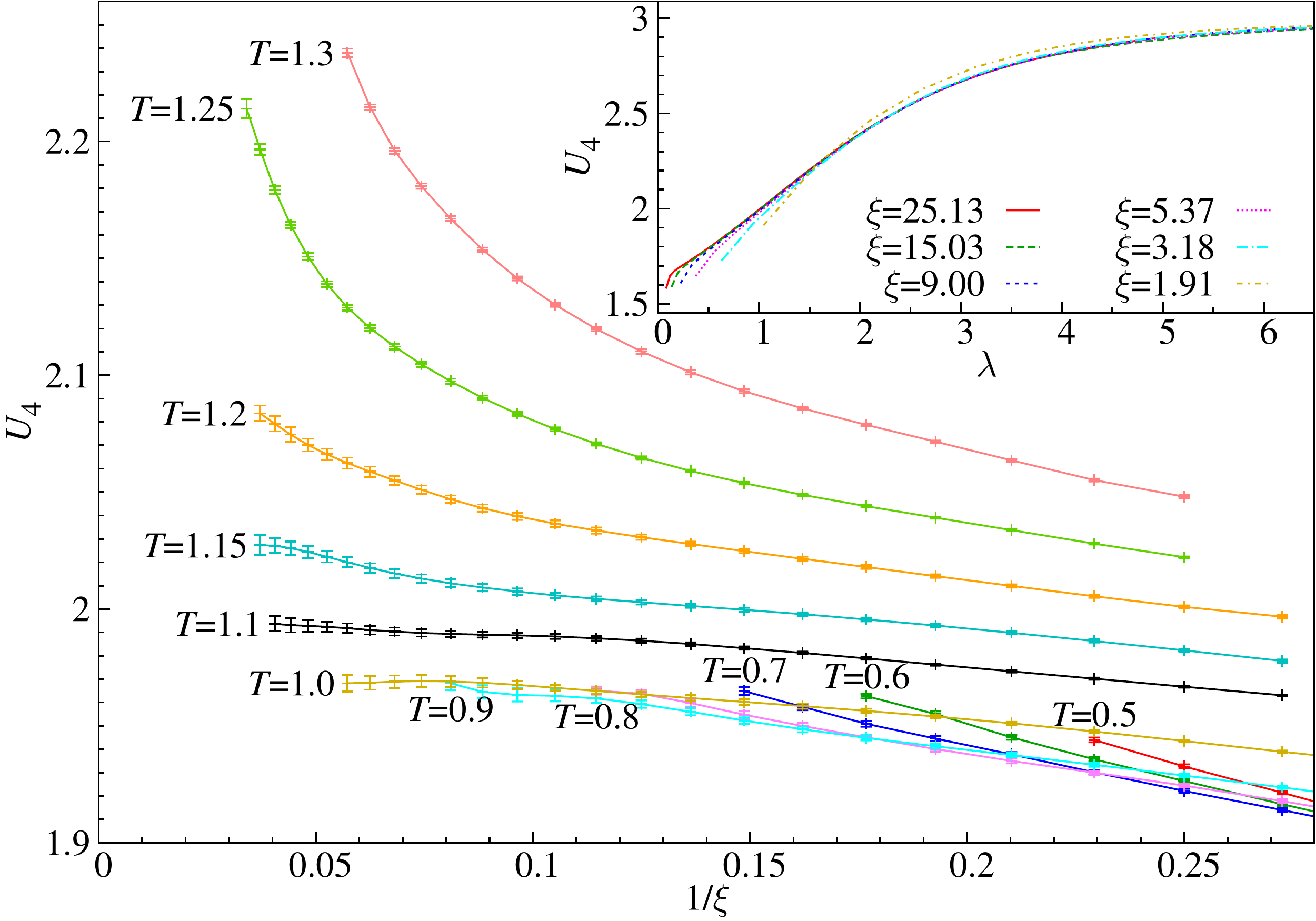}
\caption{(Color online) Binder's ratio as a function of
  $[\xi(t_\mathrm{w},T)]^{-1}$, computed for fixed dimensionless box-size
  $\lambda=1$ and several temperatures [recall that $\lambda=
    l/\xi(t_\mathrm{w})$]. {\bf Inset:} Critical Binder's ratio as a
  function of the dimensionless box-size for several
  $\xi(t_\mathrm{w})$ ($T=1.1\approx T_\mathrm{c}$). As expected by
  plugging $y=0$ in Eq.~\eqref{eq:fts}, the curve is scale-invariant
  when the small $\xi$ corrections fade away.}
\label{fig:U4}
\vspace{-0.7cm}
\end{center}
\end{figure}

Fig.~\ref{fig:U4} contains a qualitative discussion of
Eq.~\eqref{eq:fts}.  In the inset, we show data at $y=0$
(i.e. $T=1.1$, an excellent approximation to
$T_\mathrm{c}$~\cite{janus:13}). For large $\xi(t_\mathrm{w})$, $U_4$
converges to the scaling function $f(0,\lambda)$. On the other hand,
in Fig.~\ref{fig:U4}--main we show that Eq.~\eqref{eq:fts} actually
describes a cross-over in temperature. Let us fix $\lambda=1$ and
$T>T_\mathrm{c}$.  Then, $y$ becomes large and positive as
$\xi(t_\mathrm{w},T)$ grows.  We see that $U_4$ approach a
high-temperature limit (a $\lambda$-dependent renormalized coupling
constant~\cite{parisi:88}). At $T_\mathrm{c}$ we have the critical
limit because $y=0$ no matter how large $\xi(t_\mathrm{w})$ is. In the
spin-glass phase, $y$ becomes large and negative. For large $\xi$ we
reach a low-temperature limit, that has been much debated in the
past~\cite{marinari:98c,newman:98}.

\begin{figure}
\begin{center}
\includegraphics[angle=270,width=\columnwidth]{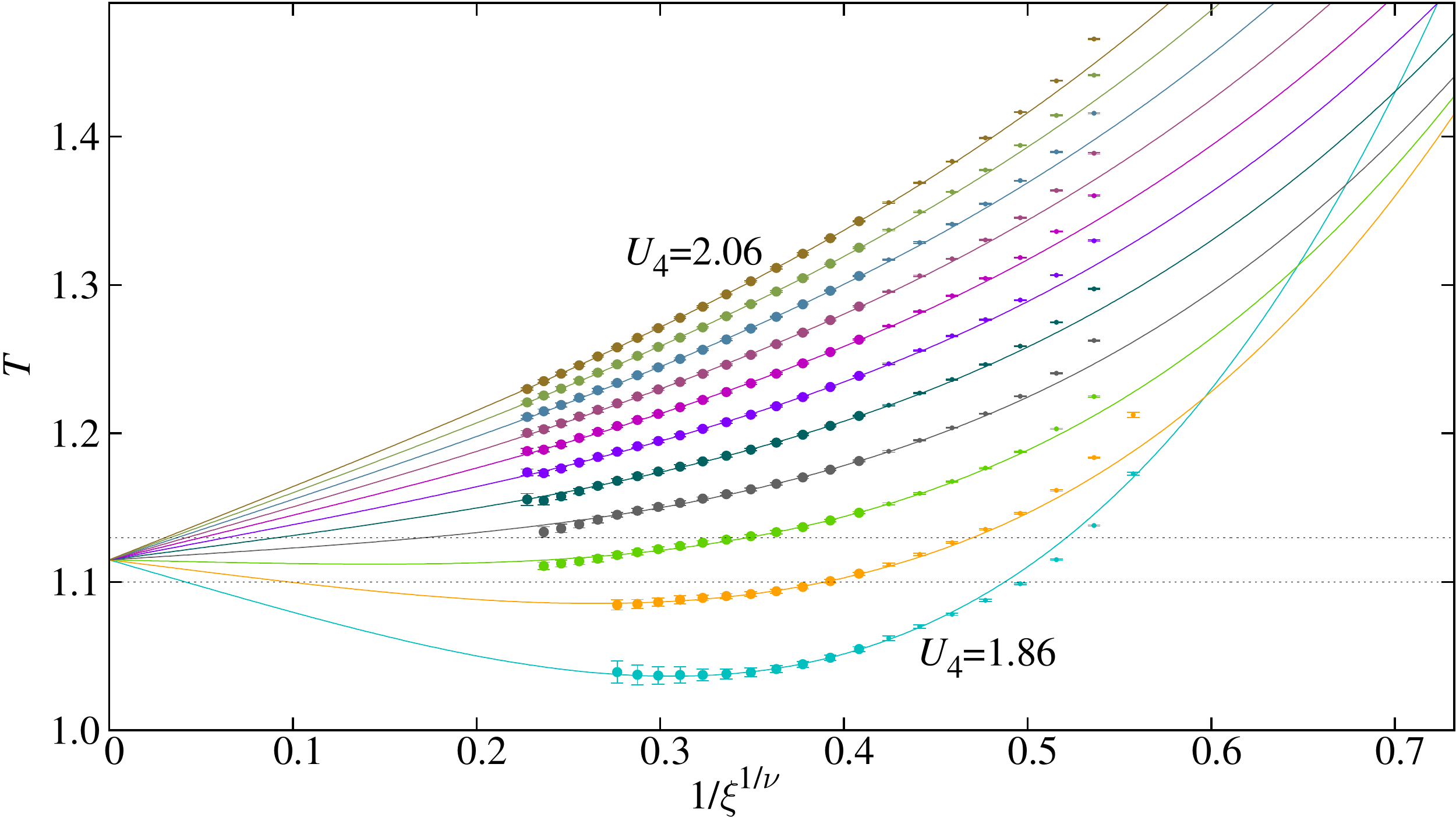}
\caption{(Color online) Joint fit to Eq.~\eqref{eq:Tc-fit} for
  $\lambda=0.75$. The $U_4$ spacing is 0.02. For all fits, the
  values of $T_\mathrm{c}$, $\nu$ and $\omega$ are held common. Big
  data points were included in the fit. The horizontal dotted lines
  correspond to $T_\mathrm{c}\pm \Delta T_\mathrm{c}$ from
  Eq.~\eqref{eq:nu-Tc-fit}.}
\label{fig:Tc-fit}
\vspace{-0.7cm}
\end{center}
\end{figure}

In order to compute the critical exponents, we decided to follow the
fixed-height method~\cite{hasenbusch:08b,janus:12}. For a fixed height $h$,
and fixed $\lambda$ and $\xi(t_\mathrm{w},T)$, we seek the temperature
$T_{h,\lambda,\xi}$ such that $U_4=h$. Eq.~\ref{eq:fts} tells us that
\begin{equation}\label{eq:Tc-fit}
T_{h,\lambda,\xi}=T_\mathrm{c} + A_{h,\lambda}\xi^{-1/\nu}+B_{h,\lambda}\xi^{-(\omega+1/\nu)}\ldots\,,
\end{equation}
where $A_{h,\lambda}$ and $B_{h,\lambda}$ are scaling amplitudes and
the dots stand for higher-order corrections to scaling. We compute
$T_\mathrm{c}$, $\nu$ and $\omega$ by performing joint fits to data
for severals $\lambda$ and $h$, see Fig.~\ref{fig:Tc-fit}
(unfortunately, the fit lacks any predictive power for exponent
$\omega$, hence we shall borrow $\omega=1.12(10)$
from~\cite{janus:13}).  In order to perform these fits, we considered
a fixed grid of coherence lengths $\xi_n=2^{n/8}$.

A major problem when fitting to Eq.~\eqref{eq:Tc-fit} is that of the
singular covariance matrix (we have many data points, but only 50
independent samples). We solve it
following~\cite{janus:08b,janus:09b}: we fit taking into account only
the diagonal part of the covariance matrix. We perform a fit for each
jackknife block, and compute the final errors from the fluctuations of
these fits. We compute as well the diagonal goodness-of-fit indicator
$\chi^2_\mathrm{diagonal}$ (the sum of the squared deviations of data
from fit, in units of their statistical error). This fitting procedure
was tested in Ref.~\cite{janus:09b} and found to be reasonably stable
for $\chi^2_\mathrm{diagonal}$ as small as half the number of degrees
of freedom.

We included in our fit results for $\lambda=0.75,1,1.25$ and $1.5$.  A
crucial issue is selecting $\xi_\mathrm{min}$, the minimal $\xi$
considered in the fit. A tradeoff should be found. The larger is
$\xi_\mathrm{min}$, the smaller are the systematic errors, but the
larger becomes the statistical uncertainty. We find a stable fit for
$\xi_\mathrm{min}\geq 2^{9/4}\approx 4.75$
($\chi^2_\mathrm{diagonal}/\text{d.o.f.}=583/665$ if
$\xi_\mathrm{min}=2^{9/4}$). However, as we enlarge $\xi_\mathrm{min}$
we find that $\chi^2_\mathrm{diagonal}/\text{d.o.f}$ decreases
monotonically while the statistical error increases. We decided to
stop at the $\xi_\mathrm{min}$ such that
$\chi^2_\mathrm{diagonal}/\text{d.o.f.}\approx 0.5$ because errors
start increasing wildly at that point. This corresponds to
$\xi_\mathrm{min}= 2^{23/8}\approx 7.33$
($\chi^2_\mathrm{diagonal}/\text{d.o.f.}=229/482$). The final result
for our fit to Eq.~\eqref{eq:Tc-fit} is
\begin{equation}\label{eq:nu-Tc-fit}
T_\mathrm{c}=1.115(15)\,,\ \nu=2.2(3)\,.
\end{equation}
For comparison, recall the equilibrium results
$T_\mathrm{c}=1.1019(29)$, $\nu=2.56(4)$ and
$\omega=1.12(10)$~\cite{janus:13}.  Varying $\omega$ within the bounds
of~\cite{janus:13} produces negligible changes in the results in
Eq.~\eqref{eq:nu-Tc-fit}. It is also interesting to see what happens
fixing $\nu$ and $\omega$ in the fit to the central values
of~\cite{janus:13} ($\xi_\mathrm{min}\geq 2^{23/8}$,
$\chi^2_\mathrm{diagonal}/\text{d.o.f.}=241/483$):
\begin{equation}\label{eq:nu-fijo-Tc-fit}
T_\mathrm{c}=1.102(8)\,, 
\end{equation}
in excellent agreement with the equilibrium result.

The anomalous dimension $\eta$ can be computed by working directly at
$T=1.1\approx T_\mathrm{c}$. We select two times $t_\mathrm{w}^{(1)}$ and
$t_\mathrm{w}^{(2)}$ such that $\xi(t_\mathrm{w}^{(1)},T_\mathrm{c})=\xi$
and  $\xi(t_\mathrm{w}^{(2)},T_\mathrm{c})=2\xi$. Then the ratio of
integrals is
\begin{equation}\label{eq:fit-eta}
I_2(t_\mathrm{w}^{(2)},T_\mathrm{c})/I_2(t_\mathrm{w}^{(1)},T_\mathrm{c}) = 2^{2-\eta}+ C_I/\xi^{\omega}+\ldots
\end{equation}
The problem with Eq.~\eqref{eq:fit-eta} is that the amplitude for
scaling corrections $C_I$ seems vanishing (within errors), so one
could be afraid that we overestimate the error. Anyhow, for
$\xi_{\text{min}}=2^{7/4}\approx 3.36$ we obtain $\eta= -0.380(7)$ and
$\chi^2_\mathrm{diagonal}/\text{d.o.f.}=10.4/14$, to be compared with
$\eta=-0.3900(36)$~\cite{janus:13} (for larger $\xi_{\text{min}}$ fits
are stable but $\chi^2_\mathrm{diagonal}/\text{d.o.f.}$ drops well
below 0.5).  Changing $\omega$ within the bounds of~\cite{janus:13}
produces a negligible change. We estimate that the error induced in
$\eta$ by the uncertainty in $T_\mathrm{c}$~\cite{janus:13} is comparable
with the statistical error obtained at $T=1.1$.

\begin{figure}
\begin{center}
\includegraphics[angle=270,width=\columnwidth]{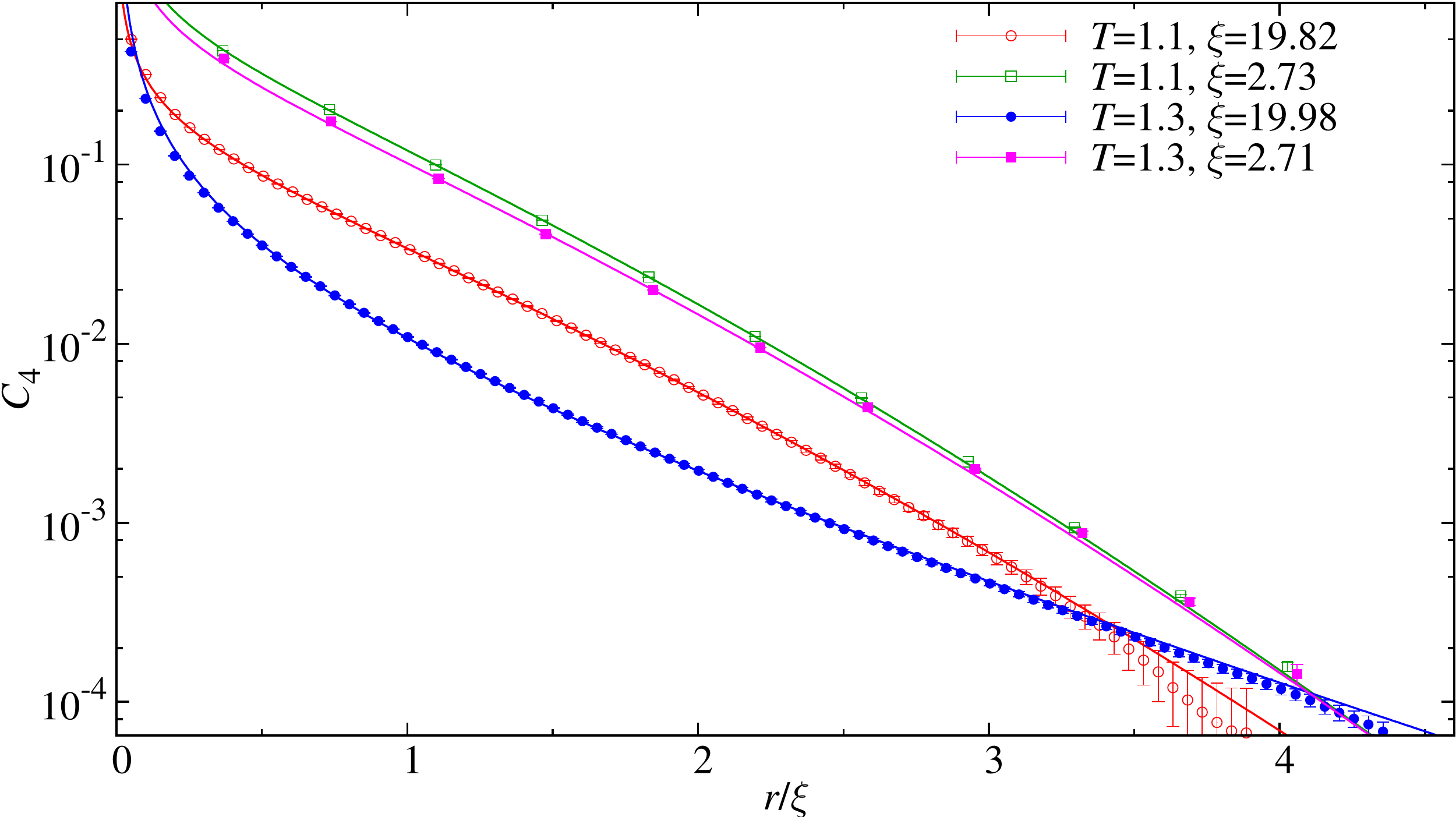}
\caption{(Color online) Temperature-dependent dynamic crossover in the
  spatial correlation function
  $C_4(r,t_{\mathrm{w}})$~\eqref{C4DEF}. We show $C_4$ vs. the
  dimensionless $r/\xi(t_\mathrm{w})$, both at $T_\mathrm{c}$ and deep
  in the paramagnetic phase ($T=1.3$). For small coherence length,
  $\xi(t_{\mathrm{w}})\approx 2.7$, data for both temperatures can be
  fit to Eq.~\eqref{eq:C4-fit} with critical parameters $b\approx 1.46$
  (this work) and $\theta\approx D-2+\eta=0.610(4)$~\cite{janus:13},
  see the continuous lines in the plot. The same parameters work for
  data at $T_{\mathrm{c}}$ and $\xi(t_\mathrm{w})\approx 20$. However,
  for such a large coherence-length, data at $T=1.3$ are better fitted
  with the three-dimensional free-field (Gaussian) parameters
  $(b=1,\theta=1)$.}
\label{fig:C4-crossover}
\vspace{-0.7cm}
\end{center}
\end{figure}

Incidentally, one may use the ratio of integrals
$I_2(t_\mathrm{w}^{(2)},T)/I_2(t_\mathrm{w}^{(1)},T)$ as a (very
noisy) substitute of the Binder's cumulant in
Eqs.~(\ref{eq:fts},\ref{eq:Tc-fit})  (see Appendix \ref{sect:I2}). In fact, one
may view the temperature crossover in Eq.~\eqref{eq:fts} as a
crossover for the $C_4(r,t_{\mathrm{w}})$ correlation
function~\eqref{C4DEF}. Indeed,
for all the
$T$ and $t_\mathrm{w}$  in this work, the functional form~\cite{marinari:96}
\begin{equation}\label{eq:C4-fit}
C_4(r,t_{\mathrm{w}})\sim \text{e}^{-[r/\xi(t_{\mathrm{w}})]^b}/r^{\theta}\,,
\end{equation}
fits satisfactorily our data. For small $y$ [i.e. at $T_\mathrm{c}$ or
  for small $\xi(t_{\mathrm{w}})$] data follows Eq.~\eqref{eq:C4-fit}
with critical parameters. However, as the coherence-length grows,
these parameters are not adequate neither for the paramagnetic phase
(at or near equilibrium, see Fig.~\ref{fig:C4-crossover}), nor for the
spin-glass phase~\cite{janus:08b,janus:09b}.

In this work, we have employed (for the first time, we believe) statics-dynamics
equivalence~\cite{janus:10b} to obtain some new physical results. In particular,
we have shown how one can study the spin-glass
transition in the dynamic regime relevant to most experiments:
non-equilibrium data on systems much larger than the coherence length.
Once we trade waiting time by coherence length, standard finite-size
scaling methods~\cite{binder:81} are very successful at describing the
temperature-dependent dynamic crossover (a real phase transition with
temperature takes place only for infinite coherence length). It is
then possible that the finite-size crossover found in
equilibrium~\cite{baityjesi:14} is the driving force behind the
apparent Universality violations found
experimentally~\cite{bouchiat:86,levy:88,petit:02,bert:04,campbell:10}.
However, an alternative explanation, logically possible but rather
dramatic, is that Universality does not hold in spin
glasses~\cite{mari:99,pleimling:05}.

Regarded as a numerical method to compute critical exponents, we note
that our thermodynamic limit approach is less accurate than
finite-size methods~\cite{hasenbusch:08b,janus:13,lulli:14b} which is
hardly a surprise.

We conclude by mentioning the two major difficulties (in our opinion)
for an analogous experimental study. On the one hand, one needs to
reach spatial resolution to study the correlation function
$C_4(r,t_\mathrm{w})$.  Progress in this direction are still
incipient. Spatial resolution has been reached only for a structural
glass~\cite{oukris:10}. For spin-glasses, recent experimental efforts
focus on confining geometries~\cite{komatsu:11,guchhait:14} (which can
be seen as an indirect way to study the correlation function). On the
other hand, the direct quench is a rather crude approximation: the
experimental sample never reaches the working temperature
instantaneously~\cite{rodriguez:03,rodriguez:13}. The protocol of
Ref.~\cite{liu:14} is, probably, more suitable to model the
experimental setup. However, as Fig.~\ref{fig:xi_tw}
shows, mixing temperatures in the dynamic evolution is a delicate
procedure that requires further investigation.

%\begin{acknowledgments}
We thank Enzo Marinari, Giorgio Parisi and Andrea Maiorano for helping
us with our first {\em Pthreads\/} programs. We also thank Andrea
Pelissetto, Giorgio Parisi and Matteo Lulli for discussing with us,
prior to publication, their very interesting and accurate
\emph{finite-time, finite-size} scaling approach~\cite{lulli:14b}.  We
thank as well Juan Jes\'us Ruiz-Lorenzo, Peter Young and David Yllanes
for discussions.

The total simulation time devoted to this project was the equivalent
of 36 days of the full (3072 AMD cores) \emph{Memento} cluster (see
\href{http://bifi.es}{http://bifi.es}).  We were partly supported by
MINECO (Spain) through research contract N$^{o}$ FIS2012-35719-C02.

%\end{acknowledgments}

\appendix

\section{Synchronous Multispin coding}\label{sect:MSC}

Modern CPUs, both Intel and AMD, support 256-bit words in their
streaming extensions. This means that one can perform basic Boolean
operations (AND, XOR, etc.)  in parallel for all the 256 bits. Now, it
is well known that the Metropolis update of a single spin can be cast
into a sequence of boolean operations, see e.g.~\cite{newman:99}. One
can use this idea to simulate several, up to 256, \emph{independent}
systems.  This approach, named asynchronous multispin coding, has been
used many times, see Refs.~\cite{leuzzi:08,banos:12,fernandez:09f,janus:13,manssen:14,lulli:14} for instance. Ref.~\cite{fang:14,feng:14}
offers a creative alternative: In their Parallel Tempering simulation
each bit represent an independent system copy (all of them evolve
under the same couplings, but at different
temperatures~\cite{hukushima:96,marinari:98b}). Instead, our aim is to
exploit the streaming extensions to speed-up the simulation of a
\emph{single} system (which is named synchronous multispin coding).

The main problem with synchronous multispin coding is that we need 256
independent random numbers, if the 256 spins coded in a word belong to
the same physical system. This breaking of parallelism is usually
regarded as a major inconvenience (see, however, Ref.~\cite{liu:14}).

For the sake of clarity, we shall first explain our geometrical set up
and then describe how one can use the Gillespie
method~\cite{gillespie:77,bortz:75} to reduce drastically the number
of needed random numbers. 

\subsection{Our multispin coding geometry}

Physical spins sit on the nodes of a $L=256$ lattice with periodic
boundary conditions. Euclidean coordinates then run as $0\leq
x,y,z\leq 255$. Each physical spin is a binary variable to be coded in
a single bit, $s_{(x,y,z)}=\pm 1$.

We pack 256 physical spins into one \emph{superspin}. Our superspins
sit in the nodes of a different lattice. It will be also a cubic
lattice with periodic boundary conditions (the overall geometry is
that of a parallelepiped, rather than a cube). The major requirement is
that nearest-neighbor spins in the physical lattice should be as well
nearest neighbors in the superspin lattice. Our solution is as
follows.

Superspins are placed at the nodes of a cubic lattice with dimensions
$L_x=L_y=L/8$, and $L_z=L/4$. The relation between physical coordinates
$(x,y,z)$ and the coordinates in the superspin lattice $(i_x,i_y,i_z)$
is
\begin{eqnarray}
x&=& b_x L_x + i_x\,,\ 0\leq i_x<L_x\,,\ 0\leq b_x < 8\,,\nonumber\\\label{eq:MSC-lattice}
y&=& b_y L_y + i_y\,,\ 0\leq i_y<L_y\,,\ 0\leq b_y < 8\,,\\
z&=& b_z L_z + i_z\,,\ 0\leq i_z<L_z\,,\ 0\leq b_z < 4\,.\nonumber
\end{eqnarray}
In this way, exactly 256 sites in the physical lattice are given the same
superspin coordinates $(i_x,i_y,i_z)$. We differenciate between them by means of
the bit index:
\begin{equation}
i_b=64 b_z+8b_y+b_x\,,\ 0\leq i_b\leq 255\,.
\end{equation}

An added bonus of Eq.~\eqref{eq:MSC-lattice} is that the parity of the
original site, namely the parity of $x+y+z$, coincides with the parity
of the corresponding superspin site $i_x+i_y+i_z$. In fact, the single
cubic lattice is bipartite. It can be regarded as two interleaved
face-centered cubic lattice. A given site is said to belong to
the~\emph{even} or the~\emph{odd} sublattice according to the parity
of $x+y+z$. For models with only nearest-neighbors interactions, sites
belonging to (say) the even sublattice interact only with the odd sites.

An important consequence of the even-odd decomposition is that it
eases parallelism. Indeed, we define the full lattice Metropolis sweep
as the update of all the $L^3/2$ even sites, followed by the update of
all the $L^3/2$ odd sites. The bipartite nature of the lattice makes it
irrelevant the updating order of sites of a given parity.  Hence,
several updating threads may legitimately concur on the same lattice,
provided that all of them simultaneously access only sites of the same parity.

\subsection{Saving random numbers}

For our synchronous multispin coding we do need to generate 256 random
numbers in order to update a single superspin. Yet, it has been
realized several times that most of the effort in generating (pseudo)
random numbers is wasted when simulating discrete models at low
temperatures~\cite{gillespie:77,bortz:75}. In fact, at a given time
the simulation may try to overcome an energy barrier $\Delta
E$. However, we should overcome it only with probability
$\text{e}^{-\Delta E/T}$.  In other words, we waste $\sim
\text{e}^{\Delta E/T}$ random numbers (that deny us the permit to
overcome the barrier) until we generate one random number that really
allows us to walk uphill in energy.  Let us plug some numbers for our
model, where the possible barrier heights are $\Delta E= 4,8$ or
$12$. So, at $T_\mathrm{c}$, in the best of cases we use only one
random number out of $\text{e}^{4/1.1}\approx 38$.

The way out is simple~\cite{gillespie:77,bortz:75}: one simulates the
random number generator. Indeed, we may regard the random-number
generator as a collection of flags. Most of the flags are red (denying
us the right to increase the energy), but there is a diluted set of
green flags (at sites where the generator does allow us to increase
the energy).  The trick is setting all flags to red by default, and
then caring only of placing green flags with the correct probability.

Before explaining how we simulate our random number generator, let us
describe it. By default, let us assume that all flags are red, for all
sites and all barriers $\Delta E= 4,8$ and $12$. Now, for each site in
the physical lattice, we draw one 64-bits uniformly distributed random
number: $0\leq R_4<1$. If $R_4<\text{e}^{-4/T}$ then we put a green
flag for $\Delta E=4$ and draw a second uniform random number $0\leq
R_8<1$. Now, if $R_8<\text{e}^{-4/T}$ we put a green flag for $\Delta
E=8$,\footnote{Probability[$R_4<\text{e}^{-4/T}$ and
  $R_8<\text{e}^{-4/T}]=\text{e}^{-8/T}$.} and draw a \emph{third}
uniform random number $0\leq R_{12}<1$. Finally, if
$R_{12}<\text{e}^{-4/T}$ we also put a green flag for $\Delta
E=12$. Of course, ours is just an instance among many valid
generators. This particular random number generator was chosen because
it is fairly easy to simulate.

Let us describe how we simulate the generation of $R_4$ (the procedure
for $R_8$ and $R_{12}$ are trivial generalizations). We generate an
integer $n_4\geq 0$, with the following meaning: One performs $n_4$
unfruitful calls to the generator, but on call $1+n_4$ we should put a
green flag. The cumulative probability for $n_4$ is
\begin{equation}
F(n_4\leq k)\equiv\text{Prob}(n_4\leq k)=1-\big(1-\text{e}^{-4/T}\big)^{k+1}\,.
\end{equation}
Hence, we just need to draw an uniform random number $0\leq R<1$ and
select $n_4=k$, where $k$ is the non-negative integer that verifies
\begin{equation}
F(k-1) \leq R <F(k)\,,\ [F(-1)\equiv -1]\,.
\end{equation}

Combining these ideas with the use of look-up-tables, we have found
that the overall cost of generating random numbers can be made quite
bearable.

\section{Interpolations}\label{sect:polynomials}

The major theme of this work is a change of variable: rather than the
the waiting time $t_\mathrm{w}$, we wish to employ the coherence
length $\xi(t_\mathrm{w})$. Besides, the quantities computed in the
l.h.s. of Eq.~(3) of the main text were obtained for a discrete set of
values of temperatures $T$, waiting times and box
sizes ($l$). However, our analysis of the r.h.s. of the same equation
assumes that the scaling variables $y$, $\lambda=l/\xi(t_\mathrm{w})$ and
$\xi(t_\mathrm{w})$ are continuous. In order to solve this problem we
perform several interpolations. 

Let us describe our interpolations.  In all cases, we perform a
jackknife error analysis. Let us stress that we are talking here about
\emph{interpolations}, rather than
extrapolations.\footnote{Exceptionally, we allowed extrapolations no
  larger than one grid-spacing in $t_\mathrm{w}$, or one fourth of the
  maximum grid-spacing in temperature.}

The easiest task is the $l$ interpolation. Data are very smooth (due
to their extreme statistical correlation) and a simple cubic spline does
an excellent job.

Let us now address $\xi(t_\mathrm{w})$. We take data for times of the
form $t_\mathrm{w}=[2^{n/4}]$ where $n$ is an integer and $[\cdots]$ is
the integer part. We find that, even for neighboring times in our
logarithmic time-mesh, the statistical fluctuations in the coherence
length are significant (see Fig.~1 in main text). However, we need a
monotonously increasing function $\xi(t_\mathrm{w})$ if we are to
invert it [that is, to obtain $t_\mathrm{w}(\xi)$]. Also it is
desiderable to have a smooth $\xi(t_\mathrm{w})$ to eliminate the
short time-scale fluctuations. Our best solution has been to fit our
data to a high-order polynomial in $\log t_\mathrm{w}$ (in the fits,
see main text, we considered only the diagonal part of the covariance
matrix). We checked that $\chi^2_\mathrm{diagonal}/\text{d.o.f}$ was
smaller than one. However, in order to avoid an excessive
data-smoothing, we enlarged the degree of the polynomial well beyond
that. Basically, we stopped before the polynomial became non
monotonically increasing in the working time-range. Notice that our
error computation (namely a different fit for each jack-knife block)
identifies spurious oscillations due to a too large-order fitting
polynomial.

Having in our hands an inverse function $t_\mathrm{w}(\xi)$ we proceed
to compute (using the same fitting approach in $\log t_\mathrm{w}$)
$U_4(\lambda,\xi,T)$. When needed, see e.g. Sect.~\ref{sect:I2}, we
interpolated in the same way the integrals $I_2(t_\mathrm{w})$.

Finally, we need to interpolate in $T$ the $U_4$ values computed at
fixed $\lambda$ and $\xi(t_\mathrm{w})$ for our simulation
temperatures.  In this case, the variations among neighboring
temperatures are typically much larger than error bars. Hence, even a
Lagrangian polynomial interpolation works well. However, when the
number of data available from the different temperatures is large, we
prefer a fit to a low-order polynomial in $T$. In practice, we restrict
ourselves to polynomials of at most fifth degree.

\section{Dynamic crossover in the correlation function}\label{sect:I2}

The dynamic cross-over (that becomes a true phase-transition with the
temperature only for infinitely long waiting time) was studied in the
main text by focusing on the four-legs correlation function of the
overlap field. One could wonder whether one could study the same
crossover on the two points correlation function.  Indeed, this was
the route chosen in Ref.~\cite{nakamura:10} (although the language in
Ref.~\cite{nakamura:10} was slightly different).

Let us start by recalling Eqs. (2,8) from the main text:\footnote{The
  standard naming \emph{two-legs} or \emph{four-legs }correlation
  function is somehow confusing in the spin-glass context. In fact,
  the product of the overlap field at two sites (the two-legs
  function) involves the product of four spins, hence the name $C_4$.}
\begin{equation}\label{eq:C4-reminder}
C_4(\boldsymbol r,t_{\mathrm{w}})\equiv\overline{\frac{1}{L^{3}}\sum_{\boldsymbol x} q_{\boldsymbol x}(t_\mathrm{w}) q_{\boldsymbol x+\boldsymbol r}(t_\mathrm{w})}\sim \frac{\text{e}^{-[r/\xi(t_{\mathrm{w}})]^b}}{r^{\theta}}\,.
\end{equation}
The asymptotic form in Eq.~\eqref{eq:C4-reminder} is expected to hold
only for $r$ much larger than the lattice spacing. Our expectations
for the asymptotic regimes are:
\begin{enumerate}
\item When we reach equilibrium in the paramagnetic phase, we expect a
  free-field behavior, namely $\theta$=1, $b=1$ in
  Eq.~\eqref{eq:C4-reminder}.
\item In the critical regime, $y$ of order one (recall from the main
  text that $y=[T-T_\mathrm{c}][\xi(t_\mathrm{w},T)]^{1/\nu}$) or
  $T=T_\mathrm{c}$, we expect $\theta=D-2+\eta$, where $D$ is the
  space dimension and $\eta$ is the anomalous dimension. We are not
  aware of any prediction for exponent $b$.  In this work, we have
  found $b=1.46(1)$.
\item There is a considerable controversy regarding the spin glass
  phase, $y\ll-1$. On the one hand, the droplets
  model~\cite{mcmillan:84,bray:87,fisher:86,fisher:88b} predicts
  $\theta=0$, although the asymptotic limit is reached fairly slowly,
  with corrections of order $1/\xi^{a\approx 0.2}$. On the other hand,
  the Replica Symmetry Breaking scenario~\cite{marinari:00} expects a
  non-trivial exponent $\theta\approx 0.37$~\cite{janus:10b} and
  corrections of order $1/\xi^\theta$. Up to our knowledge, neither of
  the two theories have predictions for exponent $b$ in
  Eq.~\eqref{eq:C4-reminder}. It was empirically found in
  Ref.~\cite{marinari:00b} that $b\approx 1.5$. In fact, we have found
  that $b=1.46(1)$ works just as well in the low temperature phase
  (see also Ref.~\cite{marinari:96}).
\end{enumerate}

In order to bypass the unknown exponent $b$, one may consider
the integrals (see ~\cite{janus:08b,janus:09b} and main text):
\begin{eqnarray}
I_n(T;t_\mathrm{w})&=&\int_0^{\infty} \mathrm{d}r\, r^n C_4(r,t_\mathrm{w},T)\,\\
&\propto&
\xi^{n+1-\theta} \int_0^{\infty} \mathrm{d}u\, u^{n-\theta} \mathrm{e}^{-u^b}\,.
\end{eqnarray}
From them we obtain the integral estimator $\xi_{12}=I_2/I_1\propto \xi$.

Our analysis will be based on the scaling properties of the integral
\begin{equation}
I_2\propto
\xi_{12}^{3-\theta}\,.
\end{equation}  
Note that, in three spatial dimensions, $\chi=4\pi I_2$, where $\chi$
is the (non-equilibrium analog of) the spin-glass
susceptibility.\footnote{The relation $\chi=4\pi I_2$ assumes spatial
  isotropy in $C_4$, which becomes an excellent approximation when
  $\xi$ grows~\cite{janus:09b}.} The analysis of
Ref.~\cite{nakamura:10} was based on the susceptibility
$\chi(T,t_\mathrm{w})$ (however, Ref.~\cite{nakamura:10} did not use
the variance reduction methods available for the computation of the
integrals $I_n$~\cite{janus:08b,janus:09b} which are most effective
because $\xi$ is much smaller than the system sizes).

\begin{figure}
\begin{center}
\includegraphics[angle=270,width=\columnwidth]{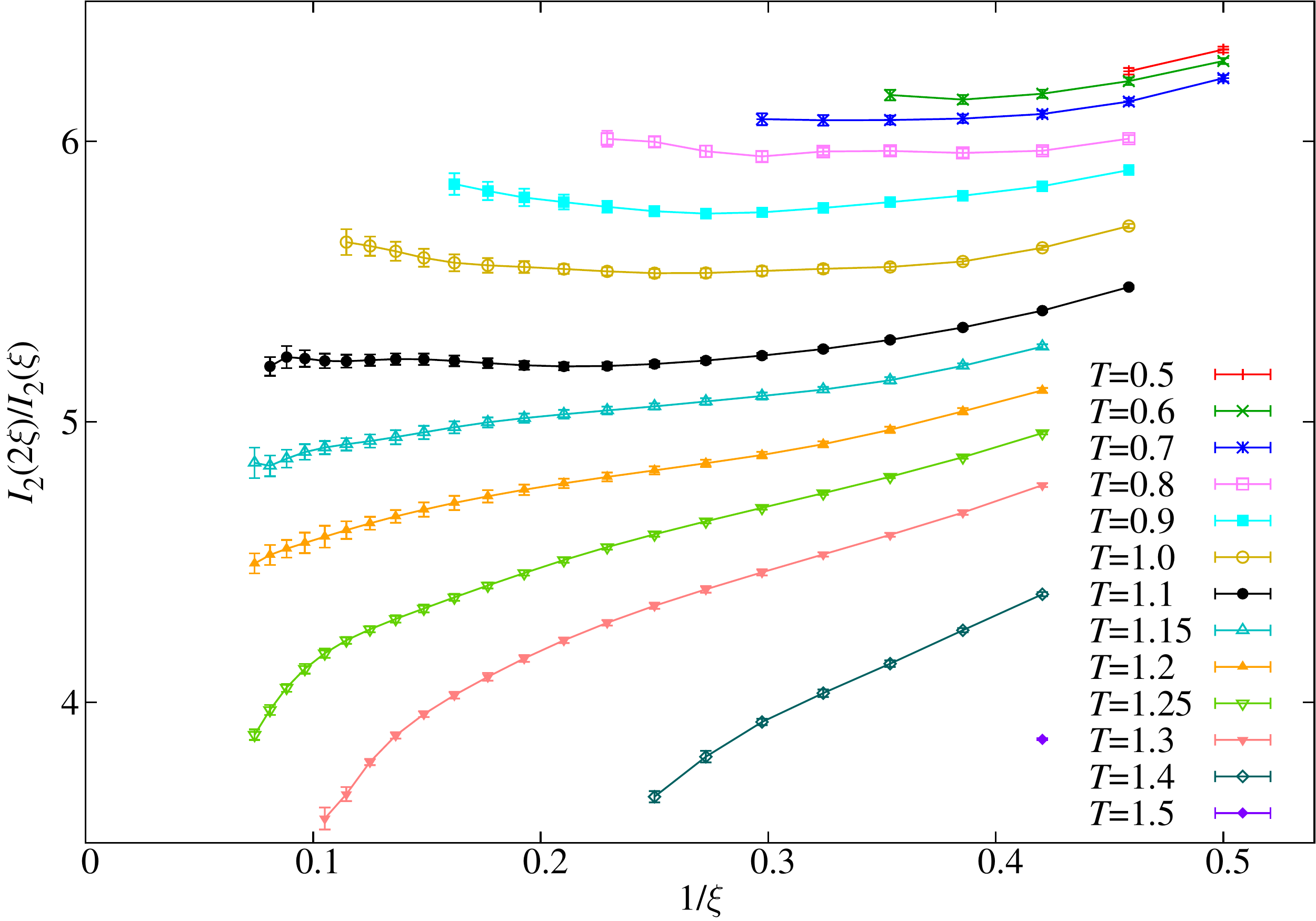}
\caption{(Color online) For several temperatures, we plot the
  susceptibility ratio of Eq.~\eqref{eq:I2-ratio} as a function of the
  inverse coherence length.}
\label{fig:I2-ratio}
\vspace{-0.7cm}
\end{center}
\end{figure}

As explained in the main text, for any given temperature we may seek
two times $t_\mathrm{w}^{(1)}$ and $t_\mathrm{w}^{(2)}$ such that and
$\xi_{12}(t_\mathrm{w}^{(2)},T)=2\xi$,
$\xi_{12}(t_\mathrm{w}^{(1)},T)=\xi$.\footnote{One could just as well
  consider pairs of times such that their coherence lengths are in any
  prescribed ratio $r$. In such case, Eq.~\eqref{eq:I2-ratio} would read as
$I_2(r\xi,T)/I_2(\xi,T)=r^{2-\eta}f_r(y)+\ldots\,.$} Hence, for $y$ of order one, we expect
\begin{equation}\label{eq:I2-ratio}
{I_2(2\xi,T)}/{I_2(\xi,T)}=2^{2-\eta} f(y)+\ldots\,,
\end{equation}
%\\
\noindent where the scaling function $f(y)$ is such that $f(y=0)=1$ and the dots
stand for corrections to scaling of order $\xi^{-\omega}$. Note that
Eq.~\eqref{eq:I2-ratio} is analogous to Eq. (3) in the main text
(where we were considering the Binder's parameter instead).\footnote{A
  statistically irrelevant artifact is the presence of wiggles in
  Fig.~\ref{fig:I2-ratio} if the order of the fitting polynomial in
  $\log t_\mathrm{w}$ is large. The origin of this wiggles has been
  known for some time~\cite{janus:08b}. The point is that each
  polynomial is evaluated twice, one in the numerator the other in the
  denominator in Eq.~\eqref{eq:I2-ratio}. In fact, the effect can be
  much alleviated by keeping the order of the polynomial limited to 13.
  Even with these polynomials, $\chi^2_\mathrm{diagonal}/\text{d.o.f}$ lies
well below one.}

The crossover implicit in Eq.~\eqref{eq:I2-ratio} is shown in
Fig.~\ref{fig:I2-ratio}, which can be directly compared with Fig. 2 in
the main text. One can consider the $\xi\to\infty$ limits in the plot:
\begin{itemize}
\item
At the critical point $T=T_\mathrm{c}$, one expects
$2^{2.3900(36)}=5.242(13)$~\cite{janus:13}. 

\item In the spin-glass phase, the droplets model predict a common
  limit $2^3=8$ for all $T<T_\mathrm{c}$, while the Replica-Symmetry
  Breaking theory expects a limit $2^{\theta_{\mathrm{RSB}}}\approx
  6.19$.

\item The paramagnetic phase is more complicated to discuss. In fact,
  for $T>T_\mathrm{c}$, the coherence length grows only up to its
  equilibrium value for that temperature, $\xi_{\text{eq}}(T)$. This
  means that all the (paramagnetic) curves in Fig.~\ref{fig:I2-ratio}
  have an end point. At this end-point, the longest time
  $t_\mathrm{w}^{(2)}$ correspond to the equilibrium regime (i.e.
  $\theta_2=1$) while the earliest time is still in the
  non-equilibrium regime. Hence, it is not easy to anticipate the
  numerical value of the paramagnetic long-time limit, obtained when
  $\xi_{\text{eq}}(T)$ tends to infinity.
\end{itemize}

Data in Fig.~\ref{fig:I2-ratio} can be analyzed in exactly the same
way we did for the Binder's parameter [see Eq.~(4) and Fig. 3 in the
  main text].  However, with the susceptibility ratio
Eq.~\eqref{eq:I2-ratio}, errors are one order of magnitude
larger. This is why we abandoned this approach.


\begin{thebibliography}{76}%
\makeatletter
\providecommand \@ifxundefined [1]{%
 \@ifx{#1\undefined}
}%
\providecommand \@ifnum [1]{%
 \ifnum #1\expandafter \@firstoftwo
 \else \expandafter \@secondoftwo
 \fi
}%
\providecommand \@ifx [1]{%
 \ifx #1\expandafter \@firstoftwo
 \else \expandafter \@secondoftwo
 \fi
}%
\providecommand \natexlab [1]{#1}%
\providecommand \enquote  [1]{``#1''}%
\providecommand \bibnamefont  [1]{#1}%
\providecommand \bibfnamefont [1]{#1}%
\providecommand \citenamefont [1]{#1}%
\providecommand \href@noop [0]{\@secondoftwo}%
\providecommand \href [0]{\begingroup \@sanitize@url \@href}%
\providecommand \@href[1]{\@@startlink{#1}\@@href}%
\providecommand \@@href[1]{\endgroup#1\@@endlink}%
\providecommand \@sanitize@url [0]{\catcode `\\12\catcode `\$12\catcode
  `\&12\catcode `\#12\catcode `\^12\catcode `\_12\catcode `\%12\relax}%
\providecommand \@@startlink[1]{}%
\providecommand \@@endlink[0]{}%
\providecommand \url  [0]{\begingroup\@sanitize@url \@url }%
\providecommand \@url [1]{\endgroup\@href {#1}{\urlprefix }}%
\providecommand \urlprefix  [0]{URL }%
\providecommand \Eprint [0]{\href }%
\providecommand \doibase [0]{http://dx.doi.org/}%
\providecommand \selectlanguage [0]{\@gobble}%
\providecommand \bibinfo  [0]{\@secondoftwo}%
\providecommand \bibfield  [0]{\@secondoftwo}%
\providecommand \translation [1]{[#1]}%
\providecommand \BibitemOpen [0]{}%
\providecommand \bibitemStop [0]{}%
\providecommand \bibitemNoStop [0]{.\EOS\space}%
\providecommand \EOS [0]{\spacefactor3000\relax}%
\providecommand \BibitemShut  [1]{\csname bibitem#1\endcsname}%
\let\auto@bib@innerbib\@empty
%</preamble>
\bibitem [{\citenamefont {Cavagna}(2009)}]{cavagna:09}%
  \BibitemOpen
  \bibfield  {author} {\bibinfo {author} {\bibfnamefont {A.}~\bibnamefont
  {Cavagna}},\ }\href@noop {} {\bibfield  {journal} {\bibinfo  {journal}
  {Physics Reports}\ }\textbf {\bibinfo {volume} {476}},\ \bibinfo {pages} {51}
  (\bibinfo {year} {2009})},\ \Eprint {http://arxiv.org/abs/arXiv:0903.4264}
  {arXiv:0903.4264} \BibitemShut {NoStop}%
\bibitem [{\citenamefont {Adam}\ and\ \citenamefont {Gibbs}(1965)}]{adam:65}%
  \BibitemOpen
  \bibfield  {author} {\bibinfo {author} {\bibfnamefont {G.}~\bibnamefont
  {Adam}}\ and\ \bibinfo {author} {\bibfnamefont {J.~H.}\ \bibnamefont
  {Gibbs}},\ }\href@noop {} {\bibfield  {journal} {\bibinfo  {journal} {J.
  Chem. Phys.}\ }\textbf {\bibinfo {volume} {43}},\ \bibinfo {pages} {139}
  (\bibinfo {year} {1965})}\BibitemShut {NoStop}%
\bibitem [{\citenamefont {Weeks}\ \emph {et~al.}(2000)\citenamefont {Weeks},
  \citenamefont {Crocker}, \citenamefont {Levitt}, \citenamefont {Schofield},\
  and\ \citenamefont {Weitz}}]{weeks:00}%
  \BibitemOpen
  \bibfield  {author} {\bibinfo {author} {\bibfnamefont {E.~R.}\ \bibnamefont
  {Weeks}}, \bibinfo {author} {\bibfnamefont {J.~C.}\ \bibnamefont {Crocker}},
  \bibinfo {author} {\bibfnamefont {A.~C.}\ \bibnamefont {Levitt}}, \bibinfo
  {author} {\bibfnamefont {A.}~\bibnamefont {Schofield}}, \ and\ \bibinfo
  {author} {\bibfnamefont {D.~A.}\ \bibnamefont {Weitz}},\ }\href@noop {}
  {\bibfield  {journal} {\bibinfo  {journal} {Science}\ }\textbf {\bibinfo
  {volume} {287}},\ \bibinfo {pages} {627} (\bibinfo {year}
  {2000})}\BibitemShut {NoStop}%
\bibitem [{\citenamefont {Berthier}\ \emph {et~al.}(2005)\citenamefont
  {Berthier}, \citenamefont {Biroli}, \citenamefont {Bouchaud}, \citenamefont
  {Cipelletti}, \citenamefont {El~Masri}, \citenamefont {{L'H{\^o}te}},
  \citenamefont {Ladieu},\ and\ \citenamefont {Pierno}}]{berthier:05}%
  \BibitemOpen
  \bibfield  {author} {\bibinfo {author} {\bibfnamefont {L.}~\bibnamefont
  {Berthier}}, \bibinfo {author} {\bibfnamefont {G.}~\bibnamefont {Biroli}},
  \bibinfo {author} {\bibfnamefont {J.-P.}\ \bibnamefont {Bouchaud}}, \bibinfo
  {author} {\bibfnamefont {L.}~\bibnamefont {Cipelletti}}, \bibinfo {author}
  {\bibfnamefont {D.}~\bibnamefont {El~Masri}}, \bibinfo {author}
  {\bibfnamefont {D.}~\bibnamefont {{L'H{\^o}te}}}, \bibinfo {author}
  {\bibfnamefont {F.}~\bibnamefont {Ladieu}}, \ and\ \bibinfo {author}
  {\bibfnamefont {M.}~\bibnamefont {Pierno}},\ }\href@noop {} {\bibfield
  {journal} {\bibinfo  {journal} {Science}\ }\textbf {\bibinfo {volume}
  {310}},\ \bibinfo {pages} {1797} (\bibinfo {year} {2005})}\BibitemShut
  {NoStop}%
\bibitem [{\citenamefont {Gutiérrez}\ \emph {et~al.}(2014)\citenamefont
  {Gutiérrez}, \citenamefont {Karmakar}, \citenamefont {Pollack},\ and\
  \citenamefont {Procaccia}}]{gutierrez:14}%
  \BibitemOpen
  \bibfield  {author} {\bibinfo {author} {\bibfnamefont {R.}~\bibnamefont
  {Gutiérrez}}, \bibinfo {author} {\bibfnamefont {S.}~\bibnamefont
  {Karmakar}}, \bibinfo {author} {\bibfnamefont {Y.~G.}\ \bibnamefont
  {Pollack}}, \ and\ \bibinfo {author} {\bibfnamefont {I.}~\bibnamefont
  {Procaccia}},\ }\href@noop {} {\  (\bibinfo {year} {2014})},\ \Eprint
  {http://arxiv.org/abs/arXiv:1409.5067} {arXiv:1409.5067} \BibitemShut
  {NoStop}%
\bibitem [{\citenamefont {Gunnarsson}\ \emph {et~al.}(1991)\citenamefont
  {Gunnarsson}, \citenamefont {Svedlindh}, \citenamefont {Nordblad},
  \citenamefont {Lundgren}, \citenamefont {Aruga},\ and\ \citenamefont
  {Ito}}]{gunnarsson:91}%
  \BibitemOpen
  \bibfield  {author} {\bibinfo {author} {\bibfnamefont {K.}~\bibnamefont
  {Gunnarsson}}, \bibinfo {author} {\bibfnamefont {P.}~\bibnamefont
  {Svedlindh}}, \bibinfo {author} {\bibfnamefont {P.}~\bibnamefont {Nordblad}},
  \bibinfo {author} {\bibfnamefont {L.}~\bibnamefont {Lundgren}}, \bibinfo
  {author} {\bibfnamefont {H.}~\bibnamefont {Aruga}}, \ and\ \bibinfo {author}
  {\bibfnamefont {A.}~\bibnamefont {Ito}},\ }\href {\doibase
  10.1103/PhysRevB.43.8199} {\bibfield  {journal} {\bibinfo  {journal} {Phys.
  Rev. B}\ }\textbf {\bibinfo {volume} {43}},\ \bibinfo {pages} {8199}
  (\bibinfo {year} {1991})}\BibitemShut {NoStop}%
\bibitem [{\citenamefont {Palassini}\ and\ \citenamefont
  {Caracciolo}(1999)}]{palassini:99}%
  \BibitemOpen
  \bibfield  {author} {\bibinfo {author} {\bibfnamefont {M.}~\bibnamefont
  {Palassini}}\ and\ \bibinfo {author} {\bibfnamefont {S.}~\bibnamefont
  {Caracciolo}},\ }\href {\doibase 10.1103/PhysRevLett.82.5128} {\bibfield
  {journal} {\bibinfo  {journal} {Phys. Rev. Lett.}\ }\textbf {\bibinfo
  {volume} {82}},\ \bibinfo {pages} {5128} (\bibinfo {year} {1999})},\ \Eprint
  {http://arxiv.org/abs/arXiv:cond-mat/9904246} {arXiv:cond-mat/9904246}
  \BibitemShut {NoStop}%
\bibitem [{\citenamefont {Ballesteros}\ \emph {et~al.}(2000)\citenamefont
  {Ballesteros}, \citenamefont {Cruz}, \citenamefont {Fernandez}, \citenamefont
  {Martin-Mayor}, \citenamefont {Pech}, \citenamefont {Ruiz-Lorenzo},
  \citenamefont {Tarancon}, \citenamefont {Tellez}, \citenamefont {Ullod},\
  and\ \citenamefont {Ungil}}]{ballesteros:00}%
  \BibitemOpen
  \bibfield  {author} {\bibinfo {author} {\bibfnamefont {H.~G.}\ \bibnamefont
  {Ballesteros}}, \bibinfo {author} {\bibfnamefont {A.}~\bibnamefont {Cruz}},
  \bibinfo {author} {\bibfnamefont {L.~A.}\ \bibnamefont {Fernandez}}, \bibinfo
  {author} {\bibfnamefont {V.}~\bibnamefont {Martin-Mayor}}, \bibinfo {author}
  {\bibfnamefont {J.}~\bibnamefont {Pech}}, \bibinfo {author} {\bibfnamefont
  {J.~J.}\ \bibnamefont {Ruiz-Lorenzo}}, \bibinfo {author} {\bibfnamefont
  {A.}~\bibnamefont {Tarancon}}, \bibinfo {author} {\bibfnamefont
  {P.}~\bibnamefont {Tellez}}, \bibinfo {author} {\bibfnamefont {C.~L.}\
  \bibnamefont {Ullod}}, \ and\ \bibinfo {author} {\bibfnamefont
  {C.}~\bibnamefont {Ungil}},\ }\href {\doibase 10.1103/PhysRevB.62.14237}
  {\bibfield  {journal} {\bibinfo  {journal} {Phys. Rev. B}\ }\textbf {\bibinfo
  {volume} {62}},\ \bibinfo {pages} {14237} (\bibinfo {year} {2000})},\ \Eprint
  {http://arxiv.org/abs/arXiv:cond-mat/0006211} {arXiv:cond-mat/0006211}
  \BibitemShut {NoStop}%
\bibitem [{\citenamefont {Vincent}\ \emph {et~al.}(1997)\citenamefont
  {Vincent}, \citenamefont {Hammann}, \citenamefont {Ocio}, \citenamefont
  {Bouchaud},\ and\ \citenamefont {Cugliandolo}}]{vincent:97}%
  \BibitemOpen
  \bibfield  {author} {\bibinfo {author} {\bibfnamefont {E.}~\bibnamefont
  {Vincent}}, \bibinfo {author} {\bibfnamefont {J.}~\bibnamefont {Hammann}},
  \bibinfo {author} {\bibfnamefont {M.}~\bibnamefont {Ocio}}, \bibinfo {author}
  {\bibfnamefont {J.-P.}\ \bibnamefont {Bouchaud}}, \ and\ \bibinfo {author}
  {\bibfnamefont {L.~F.}\ \bibnamefont {Cugliandolo}},\ }in\ \href@noop {}
  {\emph {\bibinfo {booktitle} {Complex Behavior of Glassy Systems}}},\
  \bibinfo {series and number} {\bibinfo {series} {Lecture Notes in Physics}\
  No.\ \bibinfo {number} {492}},\ \bibinfo {editor} {edited by\ \bibinfo
  {editor} {\bibfnamefont {M.}~\bibnamefont {Rub{\'{\i}}}}\ and\ \bibinfo
  {editor} {\bibfnamefont {C.}~\bibnamefont {P{\'e}rez-Vicente}}}\ (\bibinfo
  {publisher} {Springer},\ \bibinfo {year} {1997})\BibitemShut {NoStop}%
\bibitem [{\citenamefont {Bray}(1994)}]{bray:94}%
  \BibitemOpen
  \bibfield  {author} {\bibinfo {author} {\bibfnamefont {A.~J.}\ \bibnamefont
  {Bray}},\ }\href {\doibase 10.1080/00018739400101505} {\bibfield  {journal}
  {\bibinfo  {journal} {Adv. Phys.}\ }\textbf {\bibinfo {volume} {43}},\
  \bibinfo {pages} {357} (\bibinfo {year} {1994})}\BibitemShut {NoStop}%
\bibitem [{\citenamefont {Joh}\ \emph {et~al.}(1999)\citenamefont {Joh},
  \citenamefont {Orbach}, \citenamefont {Wood}, \citenamefont {Hammann},\ and\
  \citenamefont {Vincent}}]{joh:99}%
  \BibitemOpen
  \bibfield  {author} {\bibinfo {author} {\bibfnamefont {Y.~G.}\ \bibnamefont
  {Joh}}, \bibinfo {author} {\bibfnamefont {R.}~\bibnamefont {Orbach}},
  \bibinfo {author} {\bibfnamefont {G.~G.}\ \bibnamefont {Wood}}, \bibinfo
  {author} {\bibfnamefont {J.}~\bibnamefont {Hammann}}, \ and\ \bibinfo
  {author} {\bibfnamefont {E.}~\bibnamefont {Vincent}},\ }\href {\doibase
  10.1103/PhysRevLett.82.438} {\bibfield  {journal} {\bibinfo  {journal} {Phys.
  Rev. Lett.}\ }\textbf {\bibinfo {volume} {82}},\ \bibinfo {pages} {438}
  (\bibinfo {year} {1999})}\BibitemShut {NoStop}%
\bibitem [{\citenamefont {Bert}\ \emph {et~al.}(2004)\citenamefont {Bert},
  \citenamefont {Dupuis}, \citenamefont {Vincent}, \citenamefont {Hammann},\
  and\ \citenamefont {Bouchaud}}]{bert:04}%
  \BibitemOpen
  \bibfield  {author} {\bibinfo {author} {\bibfnamefont {F.}~\bibnamefont
  {Bert}}, \bibinfo {author} {\bibfnamefont {V.}~\bibnamefont {Dupuis}},
  \bibinfo {author} {\bibfnamefont {E.}~\bibnamefont {Vincent}}, \bibinfo
  {author} {\bibfnamefont {J.}~\bibnamefont {Hammann}}, \ and\ \bibinfo
  {author} {\bibfnamefont {J.-P.}\ \bibnamefont {Bouchaud}},\ }\href {\doibase
  10.1103/PhysRevLett.92.167203} {\bibfield  {journal} {\bibinfo  {journal}
  {Phys. Rev. Lett.}\ }\textbf {\bibinfo {volume} {92}},\ \bibinfo {pages}
  {167203} (\bibinfo {year} {2004})}\BibitemShut {NoStop}%
\bibitem [{\citenamefont {Rieger}(1993)}]{rieger:93}%
  \BibitemOpen
  \bibfield  {author} {\bibinfo {author} {\bibfnamefont {H.}~\bibnamefont
  {Rieger}},\ }\href {\doibase 10.1088/0305-4470/26/15/001} {\bibfield
  {journal} {\bibinfo  {journal} {J. Phys. A}\ }\textbf {\bibinfo {volume}
  {26}},\ \bibinfo {pages} {L615} (\bibinfo {year} {1993})}\BibitemShut
  {NoStop}%
\bibitem [{\citenamefont {Kisker}\ \emph {et~al.}(1996)\citenamefont {Kisker},
  \citenamefont {Santen}, \citenamefont {Schreckenberg},\ and\ \citenamefont
  {Rieger}}]{kisker:96}%
  \BibitemOpen
  \bibfield  {author} {\bibinfo {author} {\bibfnamefont {J.}~\bibnamefont
  {Kisker}}, \bibinfo {author} {\bibfnamefont {L.}~\bibnamefont {Santen}},
  \bibinfo {author} {\bibfnamefont {M.}~\bibnamefont {Schreckenberg}}, \ and\
  \bibinfo {author} {\bibfnamefont {H.}~\bibnamefont {Rieger}},\ }\href
  {\doibase 10.1103/PhysRevB.53.6418} {\bibfield  {journal} {\bibinfo
  {journal} {Phys. Rev. B}\ }\textbf {\bibinfo {volume} {53}},\ \bibinfo
  {pages} {6418} (\bibinfo {year} {1996})}\BibitemShut {NoStop}%
\bibitem [{\citenamefont {Marinari}\ \emph
  {et~al.}(2000{\natexlab{a}})\citenamefont {Marinari}, \citenamefont {Parisi},
  \citenamefont {Ricci-Tersenghi}, \citenamefont {Ruiz-Lorenzo},\ and\
  \citenamefont {Zuliani}}]{marinari:00}%
  \BibitemOpen
  \bibfield  {author} {\bibinfo {author} {\bibfnamefont {E.}~\bibnamefont
  {Marinari}}, \bibinfo {author} {\bibfnamefont {G.}~\bibnamefont {Parisi}},
  \bibinfo {author} {\bibfnamefont {F.}~\bibnamefont {Ricci-Tersenghi}},
  \bibinfo {author} {\bibfnamefont {J.~J.}\ \bibnamefont {Ruiz-Lorenzo}}, \
  and\ \bibinfo {author} {\bibfnamefont {F.}~\bibnamefont {Zuliani}},\ }\href
  {\doibase 10.1023/A:1018607809852} {\bibfield  {journal} {\bibinfo  {journal}
  {J. Stat. Phys.}\ }\textbf {\bibinfo {volume} {98}},\ \bibinfo {pages} {973}
  (\bibinfo {year} {2000}{\natexlab{a}})},\ \Eprint
  {http://arxiv.org/abs/arXiv:cond-mat/9906076} {arXiv:cond-mat/9906076}
  \BibitemShut {NoStop}%
\bibitem [{\citenamefont {Marinari}\ \emph
  {et~al.}(2000{\natexlab{b}})\citenamefont {Marinari}, \citenamefont {Parisi},
  \citenamefont {Ricci-Tersenghi},\ and\ \citenamefont
  {Ruiz-Lorenzo}}]{marinari:00b}%
  \BibitemOpen
  \bibfield  {author} {\bibinfo {author} {\bibfnamefont {E.}~\bibnamefont
  {Marinari}}, \bibinfo {author} {\bibfnamefont {G.}~\bibnamefont {Parisi}},
  \bibinfo {author} {\bibfnamefont {F.}~\bibnamefont {Ricci-Tersenghi}}, \ and\
  \bibinfo {author} {\bibfnamefont {J.~J.}\ \bibnamefont {Ruiz-Lorenzo}},\
  }\href {\doibase 10.1088/0305-4470/33/12/305} {\bibfield  {journal} {\bibinfo
   {journal} {J. Phys. A}\ }\textbf {\bibinfo {volume} {33}},\ \bibinfo {pages}
  {2373} (\bibinfo {year} {2000}{\natexlab{b}})}\BibitemShut {NoStop}%
\bibitem [{\citenamefont {Berthier}\ and\ \citenamefont
  {Bouchaud}(2002)}]{berthier:02}%
  \BibitemOpen
  \bibfield  {author} {\bibinfo {author} {\bibfnamefont {L.}~\bibnamefont
  {Berthier}}\ and\ \bibinfo {author} {\bibfnamefont {J.-P.}\ \bibnamefont
  {Bouchaud}},\ }\href {\doibase 10.1103/PhysRevB.66.054404} {\bibfield
  {journal} {\bibinfo  {journal} {Phys. Rev. B}\ }\textbf {\bibinfo {volume}
  {66}},\ \bibinfo {pages} {054404} (\bibinfo {year} {2002})}\BibitemShut
  {NoStop}%
\bibitem [{\citenamefont {Berthier}\ and\ \citenamefont
  {Young}(2005)}]{berthier:05b}%
  \BibitemOpen
  \bibfield  {author} {\bibinfo {author} {\bibfnamefont {L.}~\bibnamefont
  {Berthier}}\ and\ \bibinfo {author} {\bibfnamefont {A.~P.}\ \bibnamefont
  {Young}},\ }\href {\doibase 10.1103/PhysRevB.71.214429} {\bibfield  {journal}
  {\bibinfo  {journal} {Phys. Rev. B}\ }\textbf {\bibinfo {volume} {71}},\
  \bibinfo {pages} {214429} (\bibinfo {year} {2005})}\BibitemShut {NoStop}%
\bibitem [{\citenamefont {Jaubert}\ \emph {et~al.}(2007)\citenamefont
  {Jaubert}, \citenamefont {Chamon}, \citenamefont {Cugliandolo},\ and\
  \citenamefont {Picco}}]{jaubert:07}%
  \BibitemOpen
  \bibfield  {author} {\bibinfo {author} {\bibfnamefont {L.~C.}\ \bibnamefont
  {Jaubert}}, \bibinfo {author} {\bibfnamefont {C.}~\bibnamefont {Chamon}},
  \bibinfo {author} {\bibfnamefont {L.~F.}\ \bibnamefont {Cugliandolo}}, \ and\
  \bibinfo {author} {\bibfnamefont {M.}~\bibnamefont {Picco}},\ }\href
  {\doibase 10.1088/1742-5468/2007/05/P05001} {\bibfield  {journal} {\bibinfo
  {journal} {J. Stat. Mech.}\ }\textbf {\bibinfo {volume} {2007}},\ \bibinfo
  {pages} {P05001} (\bibinfo {year} {2007})}\BibitemShut {NoStop}%
\bibitem [{\citenamefont {Belletti}\ \emph
  {et~al.}(2008{\natexlab{a}})\citenamefont {Belletti}, \citenamefont
  {Cotallo}, \citenamefont {Cruz}, \citenamefont {Fernandez}, \citenamefont
  {Gordillo-Guerrero}, \citenamefont {Guidetti}, \citenamefont {Maiorano},
  \citenamefont {Mantovani}, \citenamefont {Marinari}, \citenamefont
  {Martin-Mayor}, \citenamefont {Sudupe}, \citenamefont {Navarro},
  \citenamefont {Parisi}, \citenamefont {Perez-Gaviro}, \citenamefont
  {Ruiz-Lorenzo}, \citenamefont {Schifano}, \citenamefont {Sciretti},
  \citenamefont {Tarancon}, \citenamefont {Tripiccione}, \citenamefont
  {Velasco},\ and\ \citenamefont {Yllanes}}]{janus:08b}%
  \BibitemOpen
  \bibfield  {author} {\bibinfo {author} {\bibfnamefont {F.}~\bibnamefont
  {Belletti}}, \bibinfo {author} {\bibfnamefont {M.}~\bibnamefont {Cotallo}},
  \bibinfo {author} {\bibfnamefont {A.}~\bibnamefont {Cruz}}, \bibinfo {author}
  {\bibfnamefont {L.~A.}\ \bibnamefont {Fernandez}}, \bibinfo {author}
  {\bibfnamefont {A.}~\bibnamefont {Gordillo-Guerrero}}, \bibinfo {author}
  {\bibfnamefont {M.}~\bibnamefont {Guidetti}}, \bibinfo {author}
  {\bibfnamefont {A.}~\bibnamefont {Maiorano}}, \bibinfo {author}
  {\bibfnamefont {F.}~\bibnamefont {Mantovani}}, \bibinfo {author}
  {\bibfnamefont {E.}~\bibnamefont {Marinari}}, \bibinfo {author}
  {\bibfnamefont {V.}~\bibnamefont {Martin-Mayor}}, \bibinfo {author}
  {\bibfnamefont {A.~M.}\ \bibnamefont {Sudupe}}, \bibinfo {author}
  {\bibfnamefont {D.}~\bibnamefont {Navarro}}, \bibinfo {author} {\bibfnamefont
  {G.}~\bibnamefont {Parisi}}, \bibinfo {author} {\bibfnamefont
  {S.}~\bibnamefont {Perez-Gaviro}}, \bibinfo {author} {\bibfnamefont {J.~J.}\
  \bibnamefont {Ruiz-Lorenzo}}, \bibinfo {author} {\bibfnamefont {S.~F.}\
  \bibnamefont {Schifano}}, \bibinfo {author} {\bibfnamefont {D.}~\bibnamefont
  {Sciretti}}, \bibinfo {author} {\bibfnamefont {A.}~\bibnamefont {Tarancon}},
  \bibinfo {author} {\bibfnamefont {R.}~\bibnamefont {Tripiccione}}, \bibinfo
  {author} {\bibfnamefont {J.~L.}\ \bibnamefont {Velasco}}, \ and\ \bibinfo
  {author} {\bibfnamefont {D.}~\bibnamefont {Yllanes}} (\bibinfo
  {collaboration} {Janus Collaboration}),\ }\href {\doibase
  10.1103/PhysRevLett.101.157201} {\bibfield  {journal} {\bibinfo  {journal}
  {Phys. Rev. Lett.}\ }\textbf {\bibinfo {volume} {101}},\ \bibinfo {pages}
  {157201} (\bibinfo {year} {2008}{\natexlab{a}})},\ \Eprint
  {http://arxiv.org/abs/arXiv:0804.1471} {arXiv:0804.1471} \BibitemShut
  {NoStop}%
\bibitem [{\citenamefont {Belletti}\ \emph {et~al.}(2009)\citenamefont
  {Belletti}, \citenamefont {Cruz}, \citenamefont {Fernandez}, \citenamefont
  {Gordillo-Guerrero}, \citenamefont {Guidetti}, \citenamefont {Maiorano},
  \citenamefont {Mantovani}, \citenamefont {Marinari}, \citenamefont
  {Martin-Mayor}, \citenamefont {Monforte}, \citenamefont {Mu{\~n}oz~Sudupe},
  \citenamefont {Navarro}, \citenamefont {Parisi}, \citenamefont
  {Perez-Gaviro}, \citenamefont {Ruiz-Lorenzo}, \citenamefont {Schifano},
  \citenamefont {Sciretti}, \citenamefont {Tarancon}, \citenamefont
  {Tripiccione},\ and\ \citenamefont {Yllanes}}]{janus:09b}%
  \BibitemOpen
  \bibfield  {author} {\bibinfo {author} {\bibfnamefont {F.}~\bibnamefont
  {Belletti}}, \bibinfo {author} {\bibfnamefont {A.}~\bibnamefont {Cruz}},
  \bibinfo {author} {\bibfnamefont {L.~A.}\ \bibnamefont {Fernandez}}, \bibinfo
  {author} {\bibfnamefont {A.}~\bibnamefont {Gordillo-Guerrero}}, \bibinfo
  {author} {\bibfnamefont {M.}~\bibnamefont {Guidetti}}, \bibinfo {author}
  {\bibfnamefont {A.}~\bibnamefont {Maiorano}}, \bibinfo {author}
  {\bibfnamefont {F.}~\bibnamefont {Mantovani}}, \bibinfo {author}
  {\bibfnamefont {E.}~\bibnamefont {Marinari}}, \bibinfo {author}
  {\bibfnamefont {V.}~\bibnamefont {Martin-Mayor}}, \bibinfo {author}
  {\bibfnamefont {J.}~\bibnamefont {Monforte}}, \bibinfo {author}
  {\bibfnamefont {A.}~\bibnamefont {Mu{\~n}oz~Sudupe}}, \bibinfo {author}
  {\bibfnamefont {D.}~\bibnamefont {Navarro}}, \bibinfo {author} {\bibfnamefont
  {G.}~\bibnamefont {Parisi}}, \bibinfo {author} {\bibfnamefont
  {S.}~\bibnamefont {Perez-Gaviro}}, \bibinfo {author} {\bibfnamefont {J.~J.}\
  \bibnamefont {Ruiz-Lorenzo}}, \bibinfo {author} {\bibfnamefont {S.~F.}\
  \bibnamefont {Schifano}}, \bibinfo {author} {\bibfnamefont {D.}~\bibnamefont
  {Sciretti}}, \bibinfo {author} {\bibfnamefont {A.}~\bibnamefont {Tarancon}},
  \bibinfo {author} {\bibfnamefont {R.}~\bibnamefont {Tripiccione}}, \ and\
  \bibinfo {author} {\bibfnamefont {D.}~\bibnamefont {Yllanes}} (\bibinfo
  {collaboration} {Janus Collaboration}),\ }\href {\doibase
  10.1007/s10955-009-9727-z} {\bibfield  {journal} {\bibinfo  {journal} {J.
  Stat. Phys.}\ }\textbf {\bibinfo {volume} {135}},\ \bibinfo {pages} {1121}
  (\bibinfo {year} {2009})},\ \Eprint {http://arxiv.org/abs/arXiv:0811.2864}
  {arXiv:0811.2864} \BibitemShut {NoStop}%
\bibitem [{\citenamefont {Manssen}\ and\ \citenamefont
  {Hartmann}(2014)}]{manssen:14}%
  \BibitemOpen
  \bibfield  {author} {\bibinfo {author} {\bibfnamefont {M.}~\bibnamefont
  {Manssen}}\ and\ \bibinfo {author} {\bibfnamefont {A.~K.}\ \bibnamefont
  {Hartmann}},\ }\href@noop {} {\  (\bibinfo {year} {2014})},\ \Eprint
  {http://arxiv.org/abs/arXiv:1411.5512} {arXiv:1411.5512} \BibitemShut
  {NoStop}%
\bibitem [{\citenamefont {Belletti}\ \emph
  {et~al.}(2008{\natexlab{b}})\citenamefont {Belletti}, \citenamefont
  {Cotallo}, \citenamefont {Cruz}, \citenamefont {Fernandez}, \citenamefont
  {Gordillo}, \citenamefont {Maiorano}, \citenamefont {Mantovani},
  \citenamefont {Marinari}, \citenamefont {Martin-Mayor}, \citenamefont
  {Mu{\~n}oz~Sudupe}, \citenamefont {Navarro}, \citenamefont {Perez-Gaviro},
  \citenamefont {Ruiz-Lorenzo}, \citenamefont {Schifano}, \citenamefont
  {Sciretti}, \citenamefont {Tarancon}, \citenamefont {Tripiccione},\ and\
  \citenamefont {Velasco}}]{janus:08}%
  \BibitemOpen
  \bibfield  {author} {\bibinfo {author} {\bibfnamefont {F.}~\bibnamefont
  {Belletti}}, \bibinfo {author} {\bibfnamefont {M.}~\bibnamefont {Cotallo}},
  \bibinfo {author} {\bibfnamefont {A.}~\bibnamefont {Cruz}}, \bibinfo {author}
  {\bibfnamefont {L.~A.}\ \bibnamefont {Fernandez}}, \bibinfo {author}
  {\bibfnamefont {A.}~\bibnamefont {Gordillo}}, \bibinfo {author}
  {\bibfnamefont {A.}~\bibnamefont {Maiorano}}, \bibinfo {author}
  {\bibfnamefont {F.}~\bibnamefont {Mantovani}}, \bibinfo {author}
  {\bibfnamefont {E.}~\bibnamefont {Marinari}}, \bibinfo {author}
  {\bibfnamefont {V.}~\bibnamefont {Martin-Mayor}}, \bibinfo {author}
  {\bibfnamefont {A.}~\bibnamefont {Mu{\~n}oz~Sudupe}}, \bibinfo {author}
  {\bibfnamefont {D.}~\bibnamefont {Navarro}}, \bibinfo {author} {\bibfnamefont
  {S.}~\bibnamefont {Perez-Gaviro}}, \bibinfo {author} {\bibfnamefont {J.~J.}\
  \bibnamefont {Ruiz-Lorenzo}}, \bibinfo {author} {\bibfnamefont {S.~F.}\
  \bibnamefont {Schifano}}, \bibinfo {author} {\bibfnamefont {D.}~\bibnamefont
  {Sciretti}}, \bibinfo {author} {\bibfnamefont {A.}~\bibnamefont {Tarancon}},
  \bibinfo {author} {\bibfnamefont {R.}~\bibnamefont {Tripiccione}}, \ and\
  \bibinfo {author} {\bibfnamefont {J.~L.}\ \bibnamefont {Velasco}} (\bibinfo
  {collaboration} {Janus Collaboration}),\ }\href {\doibase
  10.1016/j.cpc.2007.09.006} {\bibfield  {journal} {\bibinfo  {journal} {Comp.
  Phys. Comm.}\ }\textbf {\bibinfo {volume} {178}},\ \bibinfo {pages} {208}
  (\bibinfo {year} {2008}{\natexlab{b}})},\ \Eprint
  {http://arxiv.org/abs/arXiv:0704.3573} {arXiv:0704.3573} \BibitemShut
  {NoStop}%
\bibitem [{\citenamefont {Franz}\ \emph {et~al.}(1998)\citenamefont {Franz},
  \citenamefont {M{\'e}zard}, \citenamefont {Parisi},\ and\ \citenamefont
  {Peliti}}]{franz:98}%
  \BibitemOpen
  \bibfield  {author} {\bibinfo {author} {\bibfnamefont {S.}~\bibnamefont
  {Franz}}, \bibinfo {author} {\bibfnamefont {M.}~\bibnamefont {M{\'e}zard}},
  \bibinfo {author} {\bibfnamefont {G.}~\bibnamefont {Parisi}}, \ and\ \bibinfo
  {author} {\bibfnamefont {L.}~\bibnamefont {Peliti}},\ }\href {\doibase
  10.1103/PhysRevLett.81.1758} {\bibfield  {journal} {\bibinfo  {journal}
  {Phys. Rev. Lett.}\ }\textbf {\bibinfo {volume} {81}},\ \bibinfo {pages}
  {1758} (\bibinfo {year} {1998})}\BibitemShut {NoStop}%
\bibitem [{\citenamefont {Barrat}\ and\ \citenamefont
  {Berthier}(2001)}]{barrat:01}%
  \BibitemOpen
  \bibfield  {author} {\bibinfo {author} {\bibfnamefont {A.}~\bibnamefont
  {Barrat}}\ and\ \bibinfo {author} {\bibfnamefont {L.}~\bibnamefont
  {Berthier}},\ }\href {\doibase 10.1103/PhysRevLett.87.087204} {\bibfield
  {journal} {\bibinfo  {journal} {Phys. Rev. Lett.}\ }\textbf {\bibinfo
  {volume} {87}},\ \bibinfo {pages} {087204} (\bibinfo {year}
  {2001})}\BibitemShut {NoStop}%
\bibitem [{\citenamefont {Alvarez~Ba{\~n}os}\ \emph {et~al.}(2010)\citenamefont
  {Alvarez~Ba{\~n}os}, \citenamefont {Cruz}, \citenamefont {Fernandez},
  \citenamefont {Gil-Narvion}, \citenamefont {Gordillo-Guerrero}, \citenamefont
  {Guidetti}, \citenamefont {Maiorano}, \citenamefont {Mantovani},
  \citenamefont {Marinari}, \citenamefont {Martin-Mayor}, \citenamefont
  {Monforte-Garcia}, \citenamefont {Mu{\~n}oz~Sudupe}, \citenamefont {Navarro},
  \citenamefont {Parisi}, \citenamefont {Perez-Gaviro}, \citenamefont
  {Ruiz-Lorenzo}, \citenamefont {Schifano}, \citenamefont {Seoane},
  \citenamefont {Tarancon}, \citenamefont {Tripiccione},\ and\ \citenamefont
  {Yllanes}}]{janus:10b}%
  \BibitemOpen
  \bibfield  {author} {\bibinfo {author} {\bibfnamefont {R.}~\bibnamefont
  {Alvarez~Ba{\~n}os}}, \bibinfo {author} {\bibfnamefont {A.}~\bibnamefont
  {Cruz}}, \bibinfo {author} {\bibfnamefont {L.~A.}\ \bibnamefont {Fernandez}},
  \bibinfo {author} {\bibfnamefont {J.~M.}\ \bibnamefont {Gil-Narvion}},
  \bibinfo {author} {\bibfnamefont {A.}~\bibnamefont {Gordillo-Guerrero}},
  \bibinfo {author} {\bibfnamefont {M.}~\bibnamefont {Guidetti}}, \bibinfo
  {author} {\bibfnamefont {A.}~\bibnamefont {Maiorano}}, \bibinfo {author}
  {\bibfnamefont {F.}~\bibnamefont {Mantovani}}, \bibinfo {author}
  {\bibfnamefont {E.}~\bibnamefont {Marinari}}, \bibinfo {author}
  {\bibfnamefont {V.}~\bibnamefont {Martin-Mayor}}, \bibinfo {author}
  {\bibfnamefont {J.}~\bibnamefont {Monforte-Garcia}}, \bibinfo {author}
  {\bibfnamefont {A.}~\bibnamefont {Mu{\~n}oz~Sudupe}}, \bibinfo {author}
  {\bibfnamefont {D.}~\bibnamefont {Navarro}}, \bibinfo {author} {\bibfnamefont
  {G.}~\bibnamefont {Parisi}}, \bibinfo {author} {\bibfnamefont
  {S.}~\bibnamefont {Perez-Gaviro}}, \bibinfo {author} {\bibfnamefont {J.~J.}\
  \bibnamefont {Ruiz-Lorenzo}}, \bibinfo {author} {\bibfnamefont {S.~F.}\
  \bibnamefont {Schifano}}, \bibinfo {author} {\bibfnamefont {B.}~\bibnamefont
  {Seoane}}, \bibinfo {author} {\bibfnamefont {A.}~\bibnamefont {Tarancon}},
  \bibinfo {author} {\bibfnamefont {R.}~\bibnamefont {Tripiccione}}, \ and\
  \bibinfo {author} {\bibfnamefont {D.}~\bibnamefont {Yllanes}} (\bibinfo
  {collaboration} {Janus Collaboration}),\ }\href {\doibase
  10.1103/PhysRevLett.105.177202} {\bibfield  {journal} {\bibinfo  {journal}
  {Phys. Rev. Lett.}\ }\textbf {\bibinfo {volume} {105}},\ \bibinfo {pages}
  {177202} (\bibinfo {year} {2010})},\ \Eprint
  {http://arxiv.org/abs/arXiv:1003.2943} {arXiv:1003.2943} \BibitemShut
  {NoStop}%
\bibitem [{\citenamefont {J\"onsson}\ \emph {et~al.}(2002)\citenamefont
  {J\"onsson}, \citenamefont {Yoshino}, \citenamefont {Nordblad}, \citenamefont
  {Aruga~Katori},\ and\ \citenamefont {Ito}}]{jonsson:02b}%
  \BibitemOpen
  \bibfield  {author} {\bibinfo {author} {\bibfnamefont {P.~E.}\ \bibnamefont
  {J\"onsson}}, \bibinfo {author} {\bibfnamefont {H.}~\bibnamefont {Yoshino}},
  \bibinfo {author} {\bibfnamefont {P.}~\bibnamefont {Nordblad}}, \bibinfo
  {author} {\bibfnamefont {H.}~\bibnamefont {Aruga~Katori}}, \ and\ \bibinfo
  {author} {\bibfnamefont {A.}~\bibnamefont {Ito}},\ }\href {\doibase
  10.1103/PhysRevLett.88.257204} {\bibfield  {journal} {\bibinfo  {journal}
  {Phys. Rev. Lett.}\ }\textbf {\bibinfo {volume} {88}},\ \bibinfo {pages}
  {257204} (\bibinfo {year} {2002})}\BibitemShut {NoStop}%
\bibitem [{\citenamefont {Nakamae}\ \emph {et~al.}(2012)\citenamefont
  {Nakamae}, \citenamefont {Crauste-Thibierge}, \citenamefont {L'Hôte},
  \citenamefont {Vincent}, \citenamefont {Dubois}, \citenamefont {Dupuis},\
  and\ \citenamefont {Perzynski}}]{nakamae:12}%
  \BibitemOpen
  \bibfield  {author} {\bibinfo {author} {\bibfnamefont {S.}~\bibnamefont
  {Nakamae}}, \bibinfo {author} {\bibfnamefont {C.}~\bibnamefont
  {Crauste-Thibierge}}, \bibinfo {author} {\bibfnamefont {D.}~\bibnamefont
  {L'Hôte}}, \bibinfo {author} {\bibfnamefont {E.}~\bibnamefont {Vincent}},
  \bibinfo {author} {\bibfnamefont {E.}~\bibnamefont {Dubois}}, \bibinfo
  {author} {\bibfnamefont {V.}~\bibnamefont {Dupuis}}, \ and\ \bibinfo {author}
  {\bibfnamefont {R.}~\bibnamefont {Perzynski}},\ }\href {\doibase
  http://dx.doi.org/10.1063/1.4769840} {\bibfield  {journal} {\bibinfo
  {journal} {Appl. Phys. Lett.}\ }\textbf {\bibinfo {volume} {101}},\ \bibinfo
  {pages} {242409} (\bibinfo {year} {2012})}\BibitemShut {NoStop}%
\bibitem [{\citenamefont {Guchhait}\ and\ \citenamefont
  {Orbach}(2014)}]{guchhait:14}%
  \BibitemOpen
  \bibfield  {author} {\bibinfo {author} {\bibfnamefont {S.}~\bibnamefont
  {Guchhait}}\ and\ \bibinfo {author} {\bibfnamefont {R.}~\bibnamefont
  {Orbach}},\ }\href {\doibase 10.1103/PhysRevLett.112.126401} {\bibfield
  {journal} {\bibinfo  {journal} {Phys. Rev. Lett.}\ }\textbf {\bibinfo
  {volume} {112}},\ \bibinfo {pages} {126401} (\bibinfo {year}
  {2014})}\BibitemShut {NoStop}%
\bibitem [{\citenamefont {Hasenbusch}\ \emph
  {et~al.}(2008{\natexlab{a}})\citenamefont {Hasenbusch}, \citenamefont
  {Pelissetto},\ and\ \citenamefont {Vicari}}]{hasenbusch:08}%
  \BibitemOpen
  \bibfield  {author} {\bibinfo {author} {\bibfnamefont {M.}~\bibnamefont
  {Hasenbusch}}, \bibinfo {author} {\bibfnamefont {A.}~\bibnamefont
  {Pelissetto}}, \ and\ \bibinfo {author} {\bibfnamefont {E.}~\bibnamefont
  {Vicari}},\ }\href {\doibase 10.1088/1742-5468/2008/02/L02001} {\bibfield
  {journal} {\bibinfo  {journal} {J. Stat. Mech.}\ }\textbf {\bibinfo {volume}
  {2008}},\ \bibinfo {pages} {L02001} (\bibinfo {year}
  {2008}{\natexlab{a}})}\BibitemShut {NoStop}%
\bibitem [{\citenamefont {Baity-Jesi}\ \emph {et~al.}(2013)\citenamefont
  {Baity-Jesi}, \citenamefont {Ba\~{n}os}, \citenamefont {Cruz}, \citenamefont
  {Fernandez}, \citenamefont {Gil-Narvion}, \citenamefont {Gordillo-Guerrero},
  \citenamefont {Iniguez}, \citenamefont {Maiorano}, \citenamefont {Mantovani},
  \citenamefont {Marinari}, \citenamefont {Martin-Mayor}, \citenamefont
  {Monforte-Garcia}, \citenamefont {Mu{\~n}oz~Sudupe}, \citenamefont {Navarro},
  \citenamefont {Parisi}, \citenamefont {Perez-Gaviro}, \citenamefont
  {Pivanti}, \citenamefont {Ricci-Tersenghi}, \citenamefont {Ruiz-Lorenzo},
  \citenamefont {Schifano}, \citenamefont {Seoane}, \citenamefont {Tarancon},
  \citenamefont {Tripiccione},\ and\ \citenamefont {Yllanes}}]{janus:13}%
  \BibitemOpen
  \bibfield  {author} {\bibinfo {author} {\bibfnamefont {M.}~\bibnamefont
  {Baity-Jesi}}, \bibinfo {author} {\bibfnamefont {R.~A.}\ \bibnamefont
  {Ba\~{n}os}}, \bibinfo {author} {\bibfnamefont {A.}~\bibnamefont {Cruz}},
  \bibinfo {author} {\bibfnamefont {L.~A.}\ \bibnamefont {Fernandez}}, \bibinfo
  {author} {\bibfnamefont {J.~M.}\ \bibnamefont {Gil-Narvion}}, \bibinfo
  {author} {\bibfnamefont {A.}~\bibnamefont {Gordillo-Guerrero}}, \bibinfo
  {author} {\bibfnamefont {D.}~\bibnamefont {Iniguez}}, \bibinfo {author}
  {\bibfnamefont {A.}~\bibnamefont {Maiorano}}, \bibinfo {author}
  {\bibfnamefont {F.}~\bibnamefont {Mantovani}}, \bibinfo {author}
  {\bibfnamefont {E.}~\bibnamefont {Marinari}}, \bibinfo {author}
  {\bibfnamefont {V.}~\bibnamefont {Martin-Mayor}}, \bibinfo {author}
  {\bibfnamefont {J.}~\bibnamefont {Monforte-Garcia}}, \bibinfo {author}
  {\bibfnamefont {A.}~\bibnamefont {Mu{\~n}oz~Sudupe}}, \bibinfo {author}
  {\bibfnamefont {D.}~\bibnamefont {Navarro}}, \bibinfo {author} {\bibfnamefont
  {G.}~\bibnamefont {Parisi}}, \bibinfo {author} {\bibfnamefont
  {S.}~\bibnamefont {Perez-Gaviro}}, \bibinfo {author} {\bibfnamefont
  {M.}~\bibnamefont {Pivanti}}, \bibinfo {author} {\bibfnamefont
  {F.}~\bibnamefont {Ricci-Tersenghi}}, \bibinfo {author} {\bibfnamefont
  {J.~J.}\ \bibnamefont {Ruiz-Lorenzo}}, \bibinfo {author} {\bibfnamefont
  {S.~F.}\ \bibnamefont {Schifano}}, \bibinfo {author} {\bibfnamefont
  {B.}~\bibnamefont {Seoane}}, \bibinfo {author} {\bibfnamefont
  {A.}~\bibnamefont {Tarancon}}, \bibinfo {author} {\bibfnamefont
  {R.}~\bibnamefont {Tripiccione}}, \ and\ \bibinfo {author} {\bibfnamefont
  {D.}~\bibnamefont {Yllanes}} (\bibinfo {collaboration} {Janus
  Collaboration}),\ }\href {\doibase 10.1103/PhysRevB.88.224416} {\bibfield
  {journal} {\bibinfo  {journal} {Phys. Rev. B}\ }\textbf {\bibinfo {volume}
  {88}},\ \bibinfo {pages} {224416} (\bibinfo {year} {{2013}})},\ \Eprint
  {http://arxiv.org/abs/arXiv:1310.2910} {arXiv:1310.2910} \BibitemShut
  {NoStop}%
\bibitem [{\citenamefont {Ozeki}\ and\ \citenamefont {Ito}(2007)}]{ozeki:07}%
  \BibitemOpen
  \bibfield  {author} {\bibinfo {author} {\bibfnamefont {Y.}~\bibnamefont
  {Ozeki}}\ and\ \bibinfo {author} {\bibfnamefont {N.}~\bibnamefont {Ito}},\
  }\href {\doibase 10.1088/1751-8113/40/31/R01} {\bibfield  {journal} {\bibinfo
   {journal} {J. Phys. A: Math. Theor.}\ }\textbf {\bibinfo {volume} {40}},\
  \bibinfo {pages} {R149} (\bibinfo {year} {2007})}\BibitemShut {NoStop}%
\bibitem [{\citenamefont {Hohenberg}\ and\ \citenamefont
  {Halperin}(1977)}]{hohenberg:77}%
  \BibitemOpen
  \bibfield  {author} {\bibinfo {author} {\bibfnamefont {P.}~\bibnamefont
  {Hohenberg}}\ and\ \bibinfo {author} {\bibfnamefont {B.}~\bibnamefont
  {Halperin}},\ }\href {\doibase 10.1103/RevModPhys.49.435} {\bibfield
  {journal} {\bibinfo  {journal} {Rev. Mod. Phys.}\ }\textbf {\bibinfo {volume}
  {49}},\ \bibinfo {pages} {435} (\bibinfo {year} {1977})}\BibitemShut
  {NoStop}%
\bibitem [{\citenamefont {Landau}\ and\ \citenamefont
  {Binder}(2005)}]{landau:05}%
  \BibitemOpen
  \bibfield  {author} {\bibinfo {author} {\bibfnamefont {D.~P.}\ \bibnamefont
  {Landau}}\ and\ \bibinfo {author} {\bibfnamefont {K.}~\bibnamefont
  {Binder}},\ }\href@noop {} {\emph {\bibinfo {title} {A Guide to {M}onte
  {C}arlo Simulations in Statistical Physics}}},\ \bibinfo {edition} {2nd}\
  ed.\ (\bibinfo  {publisher} {Cambridge University Press},\ \bibinfo {address}
  {Cambridge},\ \bibinfo {year} {2005})\BibitemShut {NoStop}%
\bibitem [{\citenamefont {Lulli}\ \emph
  {et~al.}(2014{\natexlab{a}})\citenamefont {Lulli}, \citenamefont
  {Bernaschi},\ and\ \citenamefont {Parisi}}]{lulli:14}%
  \BibitemOpen
  \bibfield  {author} {\bibinfo {author} {\bibfnamefont {M.}~\bibnamefont
  {Lulli}}, \bibinfo {author} {\bibfnamefont {M.}~\bibnamefont {Bernaschi}}, \
  and\ \bibinfo {author} {\bibfnamefont {G.}~\bibnamefont {Parisi}},\
  }\href@noop {} {\enquote {\bibinfo {title} {Highly optimized simulations on
  single- and multi-gpu systems of 3d ising spin glass},}\ } (\bibinfo {year}
  {2014}{\natexlab{a}}),\ \bibinfo {note} {in preparation},\ \Eprint
  {http://arxiv.org/abs/arXiv:1411.0127} {arXiv:1411.0127} \BibitemShut
  {NoStop}%
\bibitem [{\citenamefont {Fang}\ \emph {et~al.}(2014)\citenamefont {Fang},
  \citenamefont {Feng}, \citenamefont {Tam}, \citenamefont {Yun}, \citenamefont
  {Moreno}, \citenamefont {Ramanujam},\ and\ \citenamefont
  {Jarrell}}]{fang:14}%
  \BibitemOpen
  \bibfield  {author} {\bibinfo {author} {\bibfnamefont {Y.}~\bibnamefont
  {Fang}}, \bibinfo {author} {\bibfnamefont {S.}~\bibnamefont {Feng}}, \bibinfo
  {author} {\bibfnamefont {K.-M.}\ \bibnamefont {Tam}}, \bibinfo {author}
  {\bibfnamefont {Z.}~\bibnamefont {Yun}}, \bibinfo {author} {\bibfnamefont
  {J.}~\bibnamefont {Moreno}}, \bibinfo {author} {\bibfnamefont
  {J.}~\bibnamefont {Ramanujam}}, \ and\ \bibinfo {author} {\bibfnamefont
  {M.}~\bibnamefont {Jarrell}},\ }\href {\doibase
  doi:10.1016/j.cpc.2014.05.020} {\bibfield  {journal} {\bibinfo  {journal}
  {Comp. Phys. Comm.}\ }\textbf {\bibinfo {volume} {185}},\ \bibinfo {pages}
  {2467–} (\bibinfo {year} {2014})},\ \Eprint
  {http://arxiv.org/abs/arXiv:1311.5582} {arXiv:1311.5582} \BibitemShut
  {NoStop}%
\bibitem [{\citenamefont {Feng}\ \emph {et~al.}(2014)\citenamefont {Feng},
  \citenamefont {Fang}, \citenamefont {Tam}, \citenamefont {Yun}, \citenamefont
  {Ramanujam}, \citenamefont {Moreno},\ and\ \citenamefont
  {Jarrell}}]{feng:14}%
  \BibitemOpen
  \bibfield  {author} {\bibinfo {author} {\bibfnamefont {S.}~\bibnamefont
  {Feng}}, \bibinfo {author} {\bibfnamefont {Y.}~\bibnamefont {Fang}}, \bibinfo
  {author} {\bibfnamefont {K.-M.}\ \bibnamefont {Tam}}, \bibinfo {author}
  {\bibfnamefont {Z.}~\bibnamefont {Yun}}, \bibinfo {author} {\bibfnamefont
  {J.}~\bibnamefont {Ramanujam}}, \bibinfo {author} {\bibfnamefont
  {J.}~\bibnamefont {Moreno}}, \ and\ \bibinfo {author} {\bibfnamefont
  {M.}~\bibnamefont {Jarrell}},\ }\href@noop {} {\  (\bibinfo {year} {2014})},\
  \Eprint {http://arxiv.org/abs/arXiv:1403.4560} {arXiv:1403.4560} \BibitemShut
  {NoStop}%
\bibitem [{\citenamefont {Gillespie}(1977)}]{gillespie:77}%
  \BibitemOpen
  \bibfield  {author} {\bibinfo {author} {\bibfnamefont {D.~T.}\ \bibnamefont
  {Gillespie}},\ }\href {\doibase 10.1021/j100540a008} {\bibfield  {journal}
  {\bibinfo  {journal} {J. Phys. Chem.}\ }\textbf {\bibinfo {volume} {81}},\
  \bibinfo {pages} {2340} (\bibinfo {year} {1977})}\BibitemShut {NoStop}%
\bibitem [{\citenamefont {Bortz}\ \emph {et~al.}(1975)\citenamefont {Bortz},
  \citenamefont {Kalos},\ and\ \citenamefont {Lebowitz}}]{bortz:75}%
  \BibitemOpen
  \bibfield  {author} {\bibinfo {author} {\bibfnamefont {A.~B.}\ \bibnamefont
  {Bortz}}, \bibinfo {author} {\bibfnamefont {M.~H.}\ \bibnamefont {Kalos}}, \
  and\ \bibinfo {author} {\bibfnamefont {J.~L.}\ \bibnamefont {Lebowitz}},\
  }\href {\doibase doi:10.1016/0021-9991(75)90060-1} {\bibfield  {journal}
  {\bibinfo  {journal} {J. Comp. Phys.}\ }\textbf {\bibinfo {volume} {17}},\
  \bibinfo {pages} {10} (\bibinfo {year} {1975})}\BibitemShut {NoStop}%
\bibitem [{\citenamefont {Bouchaud}\ \emph {et~al.}(2001)\citenamefont
  {Bouchaud}, \citenamefont {Dupuis}, \citenamefont {Hammann},\ and\
  \citenamefont {Vincent}}]{bouchaud:01}%
  \BibitemOpen
  \bibfield  {author} {\bibinfo {author} {\bibfnamefont {J.-P.}\ \bibnamefont
  {Bouchaud}}, \bibinfo {author} {\bibfnamefont {V.}~\bibnamefont {Dupuis}},
  \bibinfo {author} {\bibfnamefont {J.}~\bibnamefont {Hammann}}, \ and\
  \bibinfo {author} {\bibfnamefont {E.}~\bibnamefont {Vincent}},\ }\href
  {\doibase 10.1103/PhysRevB.65.024439} {\bibfield  {journal} {\bibinfo
  {journal} {Phys. Rev. B}\ }\textbf {\bibinfo {volume} {65}},\ \bibinfo
  {pages} {024439} (\bibinfo {year} {2001})}\BibitemShut {NoStop}%
\bibitem [{\citenamefont {Liu}\ \emph {et~al.}(2014)\citenamefont {Liu},
  \citenamefont {Polkovnikov}, \citenamefont {Sandvik},\ and\ \citenamefont
  {Young}}]{liu:14}%
  \BibitemOpen
  \bibfield  {author} {\bibinfo {author} {\bibfnamefont {C.-W.}\ \bibnamefont
  {Liu}}, \bibinfo {author} {\bibfnamefont {A.}~\bibnamefont {Polkovnikov}},
  \bibinfo {author} {\bibfnamefont {A.}~\bibnamefont {Sandvik}}, \ and\
  \bibinfo {author} {\bibfnamefont {A.~P.}\ \bibnamefont {Young}},\ }\href@noop
  {} {\  (\bibinfo {year} {2014})},\ \Eprint
  {http://arxiv.org/abs/arXiv:1411.6745} {arXiv:1411.6745} \BibitemShut
  {NoStop}%
\bibitem [{\citenamefont {Nightingale}\ and\ \citenamefont
  {Bl\"ote}(2000)}]{nightingale:00}%
  \BibitemOpen
  \bibfield  {author} {\bibinfo {author} {\bibfnamefont {M.}~\bibnamefont
  {Nightingale}}\ and\ \bibinfo {author} {\bibfnamefont {H.}~\bibnamefont
  {Bl\"ote}},\ }\href {\doibase 10.1103/PhysRevB.62.1089} {\bibfield  {journal}
  {\bibinfo  {journal} {Phys. Rev. B}\ }\textbf {\bibinfo {volume} {62}},\
  \bibinfo {pages} {1089} (\bibinfo {year} {2000})}\BibitemShut {NoStop}%
\bibitem [{\citenamefont {Mydosh}(1993)}]{mydosh:93}%
  \BibitemOpen
  \bibfield  {author} {\bibinfo {author} {\bibfnamefont {J.~A.}\ \bibnamefont
  {Mydosh}},\ }\href@noop {} {\emph {\bibinfo {title} {Spin Glasses: an
  Experimental Introduction}}}\ (\bibinfo  {publisher} {Taylor and Francis},\
  \bibinfo {address} {London},\ \bibinfo {year} {1993})\BibitemShut {NoStop}%
\bibitem [{\citenamefont {Nakamura}\ \emph {et~al.}(2003)\citenamefont
  {Nakamura}, \citenamefont {Endoh},\ and\ \citenamefont
  {Yamamoto}}]{nakamura:03}%
  \BibitemOpen
  \bibfield  {author} {\bibinfo {author} {\bibfnamefont {T.}~\bibnamefont
  {Nakamura}}, \bibinfo {author} {\bibfnamefont {S.-i.}\ \bibnamefont {Endoh}},
  \ and\ \bibinfo {author} {\bibfnamefont {T.}~\bibnamefont {Yamamoto}},\
  }\href {\doibase 10.1088/0305-4470/36/43/015} {\bibfield  {journal} {\bibinfo
   {journal} {J. Phys. A}\ }\textbf {\bibinfo {volume} {36}},\ \bibinfo {pages}
  {10895} (\bibinfo {year} {2003})}\BibitemShut {NoStop}%
\bibitem [{\citenamefont {Rom\'a}(2010)}]{roma:13}%
  \BibitemOpen
  \bibfield  {author} {\bibinfo {author} {\bibfnamefont {F.}~\bibnamefont
  {Rom\'a}},\ }\href {\doibase 10.1103/PhysRevB.82.212402} {\bibfield
  {journal} {\bibinfo  {journal} {Phys. Rev. B}\ }\textbf {\bibinfo {volume}
  {82}},\ \bibinfo {pages} {212402} (\bibinfo {year} {2010})}\BibitemShut
  {NoStop}%
\bibitem [{\citenamefont {Binder}(1981)}]{binder:81}%
  \BibitemOpen
  \bibfield  {author} {\bibinfo {author} {\bibfnamefont {K.}~\bibnamefont
  {Binder}},\ }\href {\doibase doi:10.1007/bf01293604} {\bibfield  {journal}
  {\bibinfo  {journal} {Z. Phys. B -- Condensed Matter}\ }\textbf {\bibinfo
  {volume} {43}},\ \bibinfo {pages} {119} (\bibinfo {year} {1981})}\BibitemShut
  {NoStop}%
\bibitem [{\citenamefont {Nakamura}(2010)}]{nakamura:10}%
  \BibitemOpen
  \bibfield  {author} {\bibinfo {author} {\bibfnamefont {T.}~\bibnamefont
  {Nakamura}},\ }\href {\doibase 10.1103/PhysRevB.82.014427} {\bibfield
  {journal} {\bibinfo  {journal} {Phys. Rev. B}\ }\textbf {\bibinfo {volume}
  {82}},\ \bibinfo {pages} {014427} (\bibinfo {year} {2010})}\BibitemShut
  {NoStop}%
\bibitem [{\citenamefont {Marinari}\ \emph {et~al.}(1996)\citenamefont
  {Marinari}, \citenamefont {Parisi}, \citenamefont {Ruiz-Lorenzo},\ and\
  \citenamefont {Ritort}}]{marinari:96}%
  \BibitemOpen
  \bibfield  {author} {\bibinfo {author} {\bibfnamefont {E.}~\bibnamefont
  {Marinari}}, \bibinfo {author} {\bibfnamefont {G.}~\bibnamefont {Parisi}},
  \bibinfo {author} {\bibfnamefont {J.}~\bibnamefont {Ruiz-Lorenzo}}, \ and\
  \bibinfo {author} {\bibfnamefont {F.}~\bibnamefont {Ritort}},\ }\href
  {\doibase 10.1103/PhysRevLett.76.843} {\bibfield  {journal} {\bibinfo
  {journal} {Phys. Rev. Lett.}\ }\textbf {\bibinfo {volume} {76}},\ \bibinfo
  {pages} {843} (\bibinfo {year} {1996})}\BibitemShut {NoStop}%
\bibitem [{\citenamefont {Amit}\ and\ \citenamefont
  {Martin-Mayor}(2005)}]{amit:05}%
  \BibitemOpen
  \bibfield  {author} {\bibinfo {author} {\bibfnamefont {D.~J.}\ \bibnamefont
  {Amit}}\ and\ \bibinfo {author} {\bibfnamefont {V.}~\bibnamefont
  {Martin-Mayor}},\ }\href {\doibase 10.1142/9789812775313_bmatter} {\emph
  {\bibinfo {title} {Field Theory, the Renormalization Group and Critical
  Phenomena}}},\ \bibinfo {edition} {3rd}\ ed.\ (\bibinfo  {publisher} {World
  Scientific},\ \bibinfo {address} {Singapore},\ \bibinfo {year}
  {2005})\BibitemShut {NoStop}%
\bibitem [{\citenamefont {Parisi}(1988)}]{parisi:88}%
  \BibitemOpen
  \bibfield  {author} {\bibinfo {author} {\bibfnamefont {G.}~\bibnamefont
  {Parisi}},\ }\href@noop {} {\emph {\bibinfo {title} {Statistical Field
  Theory}}}\ (\bibinfo  {publisher} {Addison-Wesley},\ \bibinfo {year}
  {1988})\BibitemShut {NoStop}%
\bibitem [{\citenamefont {Marinari}\ \emph {et~al.}(1998)\citenamefont
  {Marinari}, \citenamefont {Parisi}, \citenamefont {Ricci-Tersenghi},\ and\
  \citenamefont {Ruiz-Lorenzo}}]{marinari:98c}%
  \BibitemOpen
  \bibfield  {author} {\bibinfo {author} {\bibfnamefont {E.}~\bibnamefont
  {Marinari}}, \bibinfo {author} {\bibfnamefont {G.}~\bibnamefont {Parisi}},
  \bibinfo {author} {\bibfnamefont {F.}~\bibnamefont {Ricci-Tersenghi}}, \ and\
  \bibinfo {author} {\bibfnamefont {J.~J.}\ \bibnamefont {Ruiz-Lorenzo}},\
  }\href {\doibase 10.1088/0305-4470/31/26/001} {\bibfield  {journal} {\bibinfo
   {journal} {Journal of Physics A: Math. and Gen.}\ }\textbf {\bibinfo
  {volume} {31}},\ \bibinfo {pages} {L481} (\bibinfo {year}
  {1998})}\BibitemShut {NoStop}%
\bibitem [{\citenamefont {Newman}\ and\ \citenamefont
  {Stein}(1998)}]{newman:98}%
  \BibitemOpen
  \bibfield  {author} {\bibinfo {author} {\bibfnamefont {C.~M.}\ \bibnamefont
  {Newman}}\ and\ \bibinfo {author} {\bibfnamefont {D.~L.}\ \bibnamefont
  {Stein}},\ }\href {\doibase 10.1103/PhysRevE.57.1356} {\bibfield  {journal}
  {\bibinfo  {journal} {Phys. Rev. E}\ }\textbf {\bibinfo {volume} {57}},\
  \bibinfo {pages} {1356} (\bibinfo {year} {1998})}\BibitemShut {NoStop}%
\bibitem [{\citenamefont {Hasenbusch}\ \emph
  {et~al.}(2008{\natexlab{b}})\citenamefont {Hasenbusch}, \citenamefont
  {Pelissetto},\ and\ \citenamefont {Vicari}}]{hasenbusch:08b}%
  \BibitemOpen
  \bibfield  {author} {\bibinfo {author} {\bibfnamefont {M.}~\bibnamefont
  {Hasenbusch}}, \bibinfo {author} {\bibfnamefont {A.}~\bibnamefont
  {Pelissetto}}, \ and\ \bibinfo {author} {\bibfnamefont {E.}~\bibnamefont
  {Vicari}},\ }\href {\doibase 10.1103/PhysRevB.78.214205} {\bibfield
  {journal} {\bibinfo  {journal} {Phys. Rev. B}\ }\textbf {\bibinfo {volume}
  {78}},\ \bibinfo {pages} {214205} (\bibinfo {year}
  {2008}{\natexlab{b}})}\BibitemShut {NoStop}%
\bibitem [{\citenamefont {Ba\~{n}os}\ \emph
  {et~al.}(2012{\natexlab{a}})\citenamefont {Ba\~{n}os}, \citenamefont {Cruz},
  \citenamefont {Fernandez}, \citenamefont {Gil-Narvion}, \citenamefont
  {Gordillo-Guerrero}, \citenamefont {Guidetti}, \citenamefont {Iniguez},
  \citenamefont {Maiorano}, \citenamefont {Marinari}, \citenamefont
  {Martin-Mayor}, \citenamefont {Monforte-Garcia}, \citenamefont
  {Mu{\~n}oz~Sudupe}, \citenamefont {Navarro}, \citenamefont {Parisi},
  \citenamefont {Perez-Gaviro}, \citenamefont {Ruiz-Lorenzo}, \citenamefont
  {Schifano}, \citenamefont {Seoane}, \citenamefont {Tarancon}, \citenamefont
  {Tellez}, \citenamefont {Tripiccione},\ and\ \citenamefont
  {Yllanes}}]{janus:12}%
  \BibitemOpen
  \bibfield  {author} {\bibinfo {author} {\bibfnamefont {R.~A.}\ \bibnamefont
  {Ba\~{n}os}}, \bibinfo {author} {\bibfnamefont {A.}~\bibnamefont {Cruz}},
  \bibinfo {author} {\bibfnamefont {L.~A.}\ \bibnamefont {Fernandez}}, \bibinfo
  {author} {\bibfnamefont {J.~M.}\ \bibnamefont {Gil-Narvion}}, \bibinfo
  {author} {\bibfnamefont {A.}~\bibnamefont {Gordillo-Guerrero}}, \bibinfo
  {author} {\bibfnamefont {M.}~\bibnamefont {Guidetti}}, \bibinfo {author}
  {\bibfnamefont {D.}~\bibnamefont {Iniguez}}, \bibinfo {author} {\bibfnamefont
  {A.}~\bibnamefont {Maiorano}}, \bibinfo {author} {\bibfnamefont
  {E.}~\bibnamefont {Marinari}}, \bibinfo {author} {\bibfnamefont
  {V.}~\bibnamefont {Martin-Mayor}}, \bibinfo {author} {\bibfnamefont
  {J.}~\bibnamefont {Monforte-Garcia}}, \bibinfo {author} {\bibfnamefont
  {A.}~\bibnamefont {Mu{\~n}oz~Sudupe}}, \bibinfo {author} {\bibfnamefont
  {D.}~\bibnamefont {Navarro}}, \bibinfo {author} {\bibfnamefont
  {G.}~\bibnamefont {Parisi}}, \bibinfo {author} {\bibfnamefont
  {S.}~\bibnamefont {Perez-Gaviro}}, \bibinfo {author} {\bibfnamefont {J.~J.}\
  \bibnamefont {Ruiz-Lorenzo}}, \bibinfo {author} {\bibfnamefont {S.~F.}\
  \bibnamefont {Schifano}}, \bibinfo {author} {\bibfnamefont {B.}~\bibnamefont
  {Seoane}}, \bibinfo {author} {\bibfnamefont {A.}~\bibnamefont {Tarancon}},
  \bibinfo {author} {\bibfnamefont {P.}~\bibnamefont {Tellez}}, \bibinfo
  {author} {\bibfnamefont {R.}~\bibnamefont {Tripiccione}}, \ and\ \bibinfo
  {author} {\bibfnamefont {D.}~\bibnamefont {Yllanes}},\ }\href {\doibase
  10.1073/pnas.1203295109} {\bibfield  {journal} {\bibinfo  {journal} {Proc.
  Natl. Acad. Sci. USA}\ }\textbf {\bibinfo {volume} {{109}}},\ \bibinfo
  {pages} {6452} (\bibinfo {year} {{2012}}{\natexlab{a}})}\BibitemShut
  {NoStop}%
\bibitem [{\citenamefont {Baity-Jesi}\ \emph {et~al.}(2014)\citenamefont
  {Baity-Jesi}, \citenamefont {Fernandez}, \citenamefont {Martin-Mayor},\ and\
  \citenamefont {Sanz}}]{baityjesi:14}%
  \BibitemOpen
  \bibfield  {author} {\bibinfo {author} {\bibfnamefont {M.}~\bibnamefont
  {Baity-Jesi}}, \bibinfo {author} {\bibfnamefont {L.~A.}\ \bibnamefont
  {Fernandez}}, \bibinfo {author} {\bibfnamefont {V.}~\bibnamefont
  {Martin-Mayor}}, \ and\ \bibinfo {author} {\bibfnamefont {J.~M.}\
  \bibnamefont {Sanz}},\ }\href {\doibase 10.1103/PhysRevB.89.014202}
  {\bibfield  {journal} {\bibinfo  {journal} {Phys. Rev.}\ }\textbf {\bibinfo
  {volume} {89}},\ \bibinfo {pages} {014202} (\bibinfo {year} {2014})},\
  \Eprint {http://arxiv.org/abs/arXiv:1309.1599} {arXiv:1309.1599} \BibitemShut
  {NoStop}%
\bibitem [{\citenamefont {Bouchiat}(1986)}]{bouchiat:86}%
  \BibitemOpen
  \bibfield  {author} {\bibinfo {author} {\bibfnamefont {H.}~\bibnamefont
  {Bouchiat}},\ }\href {\doibase 10.1051/jphys:0198600470107100} {\bibfield
  {journal} {\bibinfo  {journal} {J. Phys. France}\ }\textbf {\bibinfo {volume}
  {47}},\ \bibinfo {pages} {71} (\bibinfo {year} {1986})}\BibitemShut {NoStop}%
\bibitem [{\citenamefont {L\'evy}(1988)}]{levy:88}%
  \BibitemOpen
  \bibfield  {author} {\bibinfo {author} {\bibfnamefont {L.~P.}\ \bibnamefont
  {L\'evy}},\ }\href {\doibase 10.1103/PhysRevB.38.4963} {\bibfield  {journal}
  {\bibinfo  {journal} {Phys. Rev. B}\ }\textbf {\bibinfo {volume} {38}},\
  \bibinfo {pages} {4963} (\bibinfo {year} {1988})}\BibitemShut {NoStop}%
\bibitem [{\citenamefont {Petit}\ \emph {et~al.}(2002)\citenamefont {Petit},
  \citenamefont {Fruchter},\ and\ \citenamefont {Campbell}}]{petit:02}%
  \BibitemOpen
  \bibfield  {author} {\bibinfo {author} {\bibfnamefont {D.}~\bibnamefont
  {Petit}}, \bibinfo {author} {\bibfnamefont {L.}~\bibnamefont {Fruchter}}, \
  and\ \bibinfo {author} {\bibfnamefont {I.~A.}\ \bibnamefont {Campbell}},\
  }\href {\doibase 10.1103/PhysRevLett.88.207206} {\bibfield  {journal}
  {\bibinfo  {journal} {Phys. Rev. Lett}\ }\textbf {\bibinfo {volume} {88}},\
  \bibinfo {pages} {207206} (\bibinfo {year} {2002})},\ \Eprint
  {http://arxiv.org/abs/arXiv:cond-mat/011112} {arXiv:cond-mat/011112}
  \BibitemShut {NoStop}%
\bibitem [{\citenamefont {Campbell}\ and\ \citenamefont
  {Petit}(2010)}]{campbell:10}%
  \BibitemOpen
  \bibfield  {author} {\bibinfo {author} {\bibfnamefont {I.~A.}\ \bibnamefont
  {Campbell}}\ and\ \bibinfo {author} {\bibfnamefont {D.~C.~M.~C.}\
  \bibnamefont {Petit}},\ }\href {\doibase 10.1143/JPSJ.79.011006} {\bibfield
  {journal} {\bibinfo  {journal} {J. Phys. Soc. Jpn.}\ }\textbf {\bibinfo
  {volume} {79}},\ \bibinfo {pages} {011006} (\bibinfo {year} {2010})},\
  \Eprint {http://arxiv.org/abs/arXiv:0907.5333} {arXiv:0907.5333} \BibitemShut
  {NoStop}%
\bibitem [{\citenamefont {Mari}\ and\ \citenamefont
  {Campbell}(1999)}]{mari:99}%
  \BibitemOpen
  \bibfield  {author} {\bibinfo {author} {\bibfnamefont {P.}~\bibnamefont
  {Mari}}\ and\ \bibinfo {author} {\bibfnamefont {I.}~\bibnamefont
  {Campbell}},\ }\href {\doibase 10.1103/PhysRevE.59.2653} {\bibfield
  {journal} {\bibinfo  {journal} {Phys. Rev. E}\ }\textbf {\bibinfo {volume}
  {59}},\ \bibinfo {pages} {2653} (\bibinfo {year} {1999})}\BibitemShut
  {NoStop}%
\bibitem [{\citenamefont {Pleimling}\ and\ \citenamefont
  {Campbell}(2005)}]{pleimling:05}%
  \BibitemOpen
  \bibfield  {author} {\bibinfo {author} {\bibfnamefont {M.}~\bibnamefont
  {Pleimling}}\ and\ \bibinfo {author} {\bibfnamefont {I.}~\bibnamefont
  {Campbell}},\ }\href {\doibase 10.1103/PhysRevB.72.184429} {\bibfield
  {journal} {\bibinfo  {journal} {Phys. Rev. B}\ }\textbf {\bibinfo {volume}
  {72}},\ \bibinfo {pages} {184429} (\bibinfo {year} {2005})}\BibitemShut
  {NoStop}%
\bibitem [{\citenamefont {Lulli}\ \emph
  {et~al.}(2014{\natexlab{b}})\citenamefont {Lulli}, \citenamefont {Parisi},\
  and\ \citenamefont {Pelissetto}}]{lulli:14b}%
  \BibitemOpen
  \bibfield  {author} {\bibinfo {author} {\bibfnamefont {M.}~\bibnamefont
  {Lulli}}, \bibinfo {author} {\bibfnamefont {G.}~\bibnamefont {Parisi}}, \
  and\ \bibinfo {author} {\bibfnamefont {A.}~\bibnamefont {Pelissetto}},\
  }\href@noop {} {\enquote {\bibinfo {title} {Out-of-equilibrium measure of
  critical parameters for second-order phase transitions},}\ } (\bibinfo {year}
  {2014}{\natexlab{b}}),\ \bibinfo {note} {in preparation}\BibitemShut
  {NoStop}%
\bibitem [{\citenamefont {Oukris}\ and\ \citenamefont
  {Israeloff}(2010)}]{oukris:10}%
  \BibitemOpen
  \bibfield  {author} {\bibinfo {author} {\bibfnamefont {H.}~\bibnamefont
  {Oukris}}\ and\ \bibinfo {author} {\bibfnamefont {N.~E.}\ \bibnamefont
  {Israeloff}},\ }\href {\doibase 10.1038/nphys1482} {\bibfield  {journal}
  {\bibinfo  {journal} {Nature Physics}\ }\textbf {\bibinfo {volume} {06}},\
  \bibinfo {pages} {135} (\bibinfo {year} {2010})}\BibitemShut {NoStop}%
\bibitem [{\citenamefont {Komatsu}\ \emph {et~al.}(2011)\citenamefont
  {Komatsu}, \citenamefont {L'H{\^o}te}, \citenamefont {Nakamae}, \citenamefont
  {Mosser}, \citenamefont {Konczykowski}, \citenamefont {Dubois}, \citenamefont
  {Dupuis},\ and\ \citenamefont {Perzynski}}]{komatsu:11}%
  \BibitemOpen
  \bibfield  {author} {\bibinfo {author} {\bibfnamefont {K.}~\bibnamefont
  {Komatsu}}, \bibinfo {author} {\bibfnamefont {D.}~\bibnamefont {L'H{\^o}te}},
  \bibinfo {author} {\bibfnamefont {S.}~\bibnamefont {Nakamae}}, \bibinfo
  {author} {\bibfnamefont {V.}~\bibnamefont {Mosser}}, \bibinfo {author}
  {\bibfnamefont {M.}~\bibnamefont {Konczykowski}}, \bibinfo {author}
  {\bibfnamefont {E.}~\bibnamefont {Dubois}}, \bibinfo {author} {\bibfnamefont
  {V.}~\bibnamefont {Dupuis}}, \ and\ \bibinfo {author} {\bibfnamefont
  {R.}~\bibnamefont {Perzynski}},\ }\href {\doibase
  10.1103/PhysRevLett.106.150603} {\bibfield  {journal} {\bibinfo  {journal}
  {Phys. Rev. Lett.}\ }\textbf {\bibinfo {volume} {106}},\ \bibinfo {pages}
  {150603} (\bibinfo {year} {2011})},\ \Eprint
  {http://arxiv.org/abs/arXiv:1010.4012} {arXiv:1010.4012} \BibitemShut
  {NoStop}%
\bibitem [{\citenamefont {Rodriguez}\ \emph {et~al.}(2003)\citenamefont
  {Rodriguez}, \citenamefont {Kenning},\ and\ \citenamefont
  {Orbach}}]{rodriguez:03}%
  \BibitemOpen
  \bibfield  {author} {\bibinfo {author} {\bibfnamefont {G.~F.}\ \bibnamefont
  {Rodriguez}}, \bibinfo {author} {\bibfnamefont {G.~G.}\ \bibnamefont
  {Kenning}}, \ and\ \bibinfo {author} {\bibfnamefont {R.}~\bibnamefont
  {Orbach}},\ }\href {\doibase 10.1103/PhysRevLett.91.037203} {\bibfield
  {journal} {\bibinfo  {journal} {Phys. Rev. Lett.}\ }\textbf {\bibinfo
  {volume} {91}},\ \bibinfo {pages} {037203} (\bibinfo {year}
  {2003})}\BibitemShut {NoStop}%
\bibitem [{\citenamefont {Rodriguez}\ \emph {et~al.}(2013)\citenamefont
  {Rodriguez}, \citenamefont {Kenning},\ and\ \citenamefont
  {Orbach}}]{rodriguez:13}%
  \BibitemOpen
  \bibfield  {author} {\bibinfo {author} {\bibfnamefont {G.}~\bibnamefont
  {Rodriguez}}, \bibinfo {author} {\bibfnamefont {G.}~\bibnamefont {Kenning}},
  \ and\ \bibinfo {author} {\bibfnamefont {R.}~\bibnamefont {Orbach}},\ }\href
  {\doibase 10.1103/PhysRevB.88.054302} {\bibfield  {journal} {\bibinfo
  {journal} {Phys. Rev. B}\ }\textbf {\bibinfo {volume} {88}},\ \bibinfo
  {pages} {054302} (\bibinfo {year} {2013})}\BibitemShut {NoStop}%
\bibitem [{\citenamefont {Newman}\ and\ \citenamefont
  {Barkema}(1999)}]{newman:99}%
  \BibitemOpen
  \bibfield  {author} {\bibinfo {author} {\bibfnamefont {M.~E.~J.}\
  \bibnamefont {Newman}}\ and\ \bibinfo {author} {\bibfnamefont {G.~T.}\
  \bibnamefont {Barkema}},\ }\href@noop {} {\emph {\bibinfo {title} {{M}onte
  {C}arlo Methods in Statistical Physics}}}\ (\bibinfo  {publisher} {Clarendon
  Press},\ \bibinfo {address} {Oxford},\ \bibinfo {year} {1999})\BibitemShut
  {NoStop}%
\bibitem [{\citenamefont {Leuzzi}\ \emph {et~al.}(2008)\citenamefont {Leuzzi},
  \citenamefont {Parisi}, \citenamefont {Ricci-Tersenghi},\ and\ \citenamefont
  {Ruiz-Lorenzo}}]{leuzzi:08}%
  \BibitemOpen
  \bibfield  {author} {\bibinfo {author} {\bibfnamefont {L.}~\bibnamefont
  {Leuzzi}}, \bibinfo {author} {\bibfnamefont {G.}~\bibnamefont {Parisi}},
  \bibinfo {author} {\bibfnamefont {F.}~\bibnamefont {Ricci-Tersenghi}}, \ and\
  \bibinfo {author} {\bibfnamefont {J.~J.}\ \bibnamefont {Ruiz-Lorenzo}},\
  }\href {\doibase 10.1103/PhysRevLett.101.107203} {\bibfield  {journal}
  {\bibinfo  {journal} {Phys. Rev. Lett.}\ }\textbf {\bibinfo {volume} {101}},\
  \bibinfo {pages} {107203} (\bibinfo {year} {2008})}\BibitemShut {NoStop}%
\bibitem [{\citenamefont {Ba\~{n}os}\ \emph
  {et~al.}(2012{\natexlab{b}})\citenamefont {Ba\~{n}os}, \citenamefont
  {Fernandez}, \citenamefont {Martin-Mayor},\ and\ \citenamefont
  {Young}}]{banos:12}%
  \BibitemOpen
  \bibfield  {author} {\bibinfo {author} {\bibfnamefont {R.~A.}\ \bibnamefont
  {Ba\~{n}os}}, \bibinfo {author} {\bibfnamefont {L.~A.}\ \bibnamefont
  {Fernandez}}, \bibinfo {author} {\bibfnamefont {V.}~\bibnamefont
  {Martin-Mayor}}, \ and\ \bibinfo {author} {\bibfnamefont {A.~P.}\
  \bibnamefont {Young}},\ }\href {\doibase 10.1103/PhysRevB.86.134416}
  {\bibfield  {journal} {\bibinfo  {journal} {Phys. Rev. B}\ }\textbf {\bibinfo
  {volume} {86}},\ \bibinfo {pages} {134416} (\bibinfo {year}
  {2012}{\natexlab{b}})},\ \Eprint {http://arxiv.org/abs/arXiv:1207.7014}
  {arXiv:1207.7014} \BibitemShut {NoStop}%
\bibitem [{\citenamefont {Fernandez}\ \emph {et~al.}(2010)\citenamefont
  {Fernandez}, \citenamefont {Martin-Mayor}, \citenamefont {Parisi},\ and\
  \citenamefont {Seoane}}]{fernandez:09f}%
  \BibitemOpen
  \bibfield  {author} {\bibinfo {author} {\bibfnamefont {L.~A.}\ \bibnamefont
  {Fernandez}}, \bibinfo {author} {\bibfnamefont {V.}~\bibnamefont
  {Martin-Mayor}}, \bibinfo {author} {\bibfnamefont {G.}~\bibnamefont
  {Parisi}}, \ and\ \bibinfo {author} {\bibfnamefont {B.}~\bibnamefont
  {Seoane}},\ }\href {\doibase 10.1103/PhysRevB.81.134403} {\bibfield
  {journal} {\bibinfo  {journal} {Phys. Rev. B}\ }\textbf {\bibinfo {volume}
  {81}},\ \bibinfo {pages} {134403} (\bibinfo {year} {2010})}\BibitemShut
  {NoStop}%
\bibitem [{\citenamefont {Hukushima}\ and\ \citenamefont
  {Nemoto}(1996)}]{hukushima:96}%
  \BibitemOpen
  \bibfield  {author} {\bibinfo {author} {\bibfnamefont {K.}~\bibnamefont
  {Hukushima}}\ and\ \bibinfo {author} {\bibfnamefont {K.}~\bibnamefont
  {Nemoto}},\ }\href {\doibase 10.1143/JPSJ.65.1604} {\bibfield  {journal}
  {\bibinfo  {journal} {J. Phys. Soc. Japan}\ }\textbf {\bibinfo {volume}
  {65}},\ \bibinfo {pages} {1604} (\bibinfo {year} {1996})},\ \Eprint
  {http://arxiv.org/abs/arXiv:cond-mat/9512035} {arXiv:cond-mat/9512035}
  \BibitemShut {NoStop}%
\bibitem [{\citenamefont {Marinari}(1998)}]{marinari:98b}%
  \BibitemOpen
  \bibfield  {author} {\bibinfo {author} {\bibfnamefont {E.}~\bibnamefont
  {Marinari}},\ }in\ \href@noop {} {\emph {\bibinfo {booktitle} {Advances in
  Computer Simulation}}},\ \bibinfo {editor} {edited by\ \bibinfo {editor}
  {\bibfnamefont {J.}~\bibnamefont {Kerst\'esz}}\ and\ \bibinfo {editor}
  {\bibfnamefont {I.}~\bibnamefont {Kondor}}}\ (\bibinfo  {publisher}
  {Springer-Verlag},\ \bibinfo {year} {1998})\BibitemShut {NoStop}%
\bibitem [{\citenamefont {McMillan}(1984)}]{mcmillan:84}%
  \BibitemOpen
  \bibfield  {author} {\bibinfo {author} {\bibfnamefont {W.~L.}\ \bibnamefont
  {McMillan}},\ }\href {\doibase 10.1088/0022-3719/17/18/010} {\bibfield
  {journal} {\bibinfo  {journal} {J. Phys. C: Solid State Phys.}\ }\textbf
  {\bibinfo {volume} {17}},\ \bibinfo {pages} {3179} (\bibinfo {year}
  {1984})}\BibitemShut {NoStop}%
\bibitem [{\citenamefont {Bray}\ and\ \citenamefont {Moore}(1987)}]{bray:87}%
  \BibitemOpen
  \bibfield  {author} {\bibinfo {author} {\bibfnamefont {A.~J.}\ \bibnamefont
  {Bray}}\ and\ \bibinfo {author} {\bibfnamefont {M.~A.}\ \bibnamefont
  {Moore}},\ }in\ \href@noop {} {\emph {\bibinfo {booktitle} {Heidelberg
  Colloquium on Glassy Dynamics}}},\ \bibinfo {series and number} {\bibinfo
  {series} {Lecture Notes in Physics}\ No.\ \bibinfo {number} {275}},\ \bibinfo
  {editor} {edited by\ \bibinfo {editor} {\bibfnamefont {J.~L.}\ \bibnamefont
  {van Hemmen}}\ and\ \bibinfo {editor} {\bibfnamefont {I.}~\bibnamefont
  {Morgenstern}}}\ (\bibinfo  {publisher} {Springer},\ \bibinfo {address}
  {Berlin},\ \bibinfo {year} {1987})\BibitemShut {NoStop}%
\bibitem [{\citenamefont {Fisher}\ and\ \citenamefont
  {Huse}(1986)}]{fisher:86}%
  \BibitemOpen
  \bibfield  {author} {\bibinfo {author} {\bibfnamefont {D.~S.}\ \bibnamefont
  {Fisher}}\ and\ \bibinfo {author} {\bibfnamefont {D.~A.}\ \bibnamefont
  {Huse}},\ }\href {\doibase 10.1103/PhysRevLett.56.1601} {\bibfield  {journal}
  {\bibinfo  {journal} {Phys. Rev. Lett.}\ }\textbf {\bibinfo {volume} {56}},\
  \bibinfo {pages} {1601} (\bibinfo {year} {1986})}\BibitemShut {NoStop}%
\bibitem [{\citenamefont {Fisher}\ and\ \citenamefont
  {Huse}(1988)}]{fisher:88b}%
  \BibitemOpen
  \bibfield  {author} {\bibinfo {author} {\bibfnamefont {D.~S.}\ \bibnamefont
  {Fisher}}\ and\ \bibinfo {author} {\bibfnamefont {D.~A.}\ \bibnamefont
  {Huse}},\ }\href {\doibase 10.1103/PhysRevB.38.386} {\bibfield  {journal}
  {\bibinfo  {journal} {Phys. Rev. B}\ }\textbf {\bibinfo {volume} {38}},\
  \bibinfo {pages} {386} (\bibinfo {year} {1988})}\BibitemShut {NoStop}%
\end{thebibliography}
\end{document}